\documentclass[
  aps,
  prd,
  onecolumn,
  10pt,
  a4paper,
  notitlepage,
  nofootinbib,
  floatfix
]{revtex4-2}

\usepackage[utf8]{inputenc}
\usepackage[T1]{fontenc}
\usepackage{lmodern}
\usepackage[a4paper,margin=2.2cm]{geometry}

\usepackage{amsmath}
\usepackage{amssymb}
\usepackage{mathtools}
\usepackage{amsthm}
\usepackage{bm}
\usepackage{mathrsfs}

\usepackage{graphicx}
\usepackage{booktabs}
\usepackage{graphicx}
\usepackage{booktabs}
\usepackage{microtype}
\usepackage{xcolor}

\usepackage{tikz}
\usepackage{pgfplots}
\pgfplotsset{compat=1.18}
\usetikzlibrary{arrows.meta,positioning,shadows}

\usepackage[
  colorlinks=true,
  linkcolor=blue,
  citecolor=blue,
  urlcolor=blue
]{hyperref}

\usepackage{orcidlink}

\newcommand{\dd}{\mathrm d}
\newcommand{\cH}{\mathcal H}
\newcommand{\cE}{\mathcal E}
\newcommand{\Div}{\operatorname{div}}

\theoremstyle{plain}
\newtheorem{theorem}{Theorem}
\newtheorem{proposition}[theorem]{Proposition}

\theoremstyle{definition}

\theoremstyle{remark}

\begin{document}

\title{On Chen's Extended Action for $\kappa(R,T)$ Gravity:\\
Gravity in a Polarizable Medium, Material Folds\\
and a No-Bounce Theorem}

\author{Gin\'es R. P\'erez Teruel\,
\orcidlink{0000-0001-8363-1366}}

\email{gines.landau@gmail.com}

\affiliation{Consejer\'ia de Desarrollo Educativo y Formaci\'on Profesional, Junta de Andaluc\'ia, Huelva, Spain}
\begin{abstract}
Chen's extended action, built with an adjoint field, reproduces the field equations of $\kappa(R,T)$ gravity from a local variational principle, and we show that it does physical work. In flat FLRW, Dirac reduction gives a Hamiltonian independent of $\kappa(T)$ on regular branches, where the coupling enters through a second-class matter constraint. The physical stress tensor is not generally conserved, but a dressed particle number is. An exact decomposition of the effective source motivates a polarizable-medium interpretation: $\kappa=0$ corresponds to total screening, the repulsive phase to overscreening, and a fold of the matter map to saturation. For linear coupling and dust, the expanding low-density branch has a past endpoint at the fold, where the volume is minimal and the density reaches $\rho_c/2$ at finite curvature. Dirac consistency rules out smooth Hamiltonian continuation through the fold. More generally, a flat-FLRW bounce is excluded for $\kappa>0$, positive density and $w>-1$. The linear dust model reproduces the effective Friedmann equation of loop quantum cosmology (LQC), but the two models obey different conservation laws. An effective pressure $-\rho^2/\rho_c$ restores the homogeneous LQC dynamics, and the fold density coincides with the signature-change threshold in the deformed-algebra approach to LQC. The reduced dust constraint admits an exactly solvable Koopman--von Neumann transport quantization. The adjoint field acts as a response variable, motivating an open-system interpretation.
\end{abstract}

\maketitle

\tableofcontents
\bigskip

\section{Introduction}\label{sec:intro}

In $\kappa(R,T)$ gravity \cite{Perez} the gravitational coupling of general relativity is promoted to a function of the curvature scalar and of the trace of the energy--momentum tensor,
\begin{equation}\label{eq:E}
E_{\mu\nu}[g,\chi]:=G_{\mu\nu}-\Lambda g_{\mu\nu}-\kappa(R,T)\,T_{\mu\nu}=0,\qquad T=g^{\mu\nu}T_{\mu\nu}.
\end{equation}
Here $\chi$ collectively denotes the matter variables
entering the prescribed stress tensor
$T_{\mu\nu}[g,\chi]$; no independent matter action
is assumed. The contracted Bianchi identity then replaces the conservation of $T_{\mu\nu}$ by that of $\kappa T_{\mu\nu}$. The theory is formulated directly at the level of the field equations, and was introduced as a gravitational analogue of Maxwell's displacement current, a source term added to complete the field equations \cite{Perez}. The lack of an action principle has been a recurrent objection to it: without an action, there is no canonical structure, no Hamiltonian analysis of its degrees of freedom, and no starting point for quantization.

Since its introduction, the theory has been explored in several directions. Cosmological applications include accelerated expansion and empirical parametrizations of the expansion history \cite{AhmedPradhan2020,AhmedPradhan2022,Dixit2022}, as well as the thermodynamics of the accelerating universe \cite{Dixit2023}. In the strong-field regime, stable compact stars \cite{Teruel2022,TaserDogru2023}, wormholes \cite{Sarkar2022,Singh2024} and gravastar configurations \cite{Teruel2024} have been constructed and used to test the theory. More recently, the critical regime $\kappa=0$ has been analysed \cite{Teruel2026}. There it was shown that the apparent singularity of the non-conservation equation stems from its rewritten form rather than from the field equations, and that regular critical hypersurfaces act as gravitational screening surfaces that obstruct a global invertible reconstruction of the physical matter
from the effective source. All of these works use the field equations \eqref{eq:E} directly.

The construction of a consistent variational formulation was listed as an open problem in \cite{Teruel2026}. Chen has since constructed a local extended action that reproduces \eqref{eq:E} exactly \cite{Chen}. It introduces an adjoint multiplier tensor and belongs to the established family of adjoint, or formal, Lagrangians; Section~\ref{sec:chen} reviews the action and its place in that family. The existence of such an action is therefore not specific to $\kappa(R,T)$. What matters is whether the action does useful work: whether its canonical structure is consistent, and whether it reveals properties of the theory that are not apparent at the level of \eqref{eq:E}. A second question is whether the doubled form is merely available or actually natural for this theory. We argue that it is natural, for two reasons. First, for barotropic matter an ordinary effective-fluid action is available where the constitutive reconstruction is regular. At material folds, recovering the original physical density becomes branch-dependent, without necessarily making the effective action singular. Critical and negative-coupling configurations also require care in identifying the physical matter variables.
Chen's adjoint action avoids this reconstruction:
it enforces the prescribed equations directly,
without dividing by $\kappa$ or inverting the
matter map (Section~\ref{sec:adjoint}). Second, $\kappa(R,T)$ describes matter that exchanges energy, through a time-symmetric balance law, with a medium whose stress is slaved to the matter (Section~\ref{sec:medium}). Since the medium has no dynamics of its own, the theory can be read as the effective description of a subsystem whose environment has been eliminated. For such descriptions doubled actions are not an artifice but the appropriate variational framework, as in the Schwinger--Keldysh (closed-time-path) formalism \cite{Schwinger,Keldysh}, in Galley's treatment of nonconservative systems \cite{Galley} and, for gravity, in stochastic semiclassical gravity \cite{HuVerdaguer}. In this light the adjoint field acquires a physical role, as the response variable and a candidate difference field, and the theory acquires a concrete candidate for the environment.

This note examines that question in the cosmological sector and in the vacuum quadratic theory. The main results are the following.
\begin{enumerate}
\item The flat-FLRW reduction of \eqref{eq:S} is a consistent Dirac system with one physical canonical pair. On every regular matter branch its reduced Hamiltonian is independent of the functional form of $\kappa(T)$ (Theorem~\ref{thm:universal}). The geometry therefore obeys the equations of general relativity with an effective fluid, but only branch by branch: critical surfaces $\kappa=0$ and material folds obstruct a global identification. The first obstruction, found in \cite{Teruel2026}, belongs to inhomogeneous configurations, since an FLRW trajectory with non-zero effective mass never reaches $\kappa=0$. The second, found here, belongs to cosmology.
\item Matter in $\kappa(R,T)$ is generically non-conservative, with an exchange rate $\Gamma=-\dot\kappa/\kappa$ fixed by the coupling, of the form familiar from matter-creation cosmologies \cite{Prigogine}. In non-static flat FLRW, ordinary conservation is incompatible with the modified balance law
on an evolving interval where $\kappa_T\neq0$, for $\rho>0$ and
$w\neq-1,1/3$ (Proposition~\ref{prop:conservation}). The balance law is time-symmetric, and a dressed particle number is exactly conserved (Eq.~\eqref{eq:dressed}).
\item The theory can be rewritten exactly as gravity in a polarizable medium, with an induced source of susceptibility $\chi_g=\kappa/\kappa_0-1$. The dressed number is then the sum of free and bound parts, the gravitational counterpart of free plus bound charge, and this makes the original displacement-current analogy precise. Critical surfaces, the repulsive phase and the fold appear as total screening, over-screening and saturation (Section~\ref{sec:medium}).
\item For a linear coupling, the point where the second-class structure of the action degenerates coincides with a fold of the matter map, on which no consistent Hamiltonian evolution with non-degenerate lapse can remain (Proposition~\ref{prop:dirac}). This result is specific to Chen's action: the loss of invertibility of the matter map is an obstruction to the Hamiltonian evolution, not a singularity of the variables. There the geometry is analytic and the matter density has a square-root singularity; an expanding solution has no continuation to earlier times for $\Lambda\geq0$, and the reconstruction of its matter is two-valued to the future (Proposition~\ref{prop:fold}). Expanding dust universes of the linear model on the low-density branch therefore begin at a minimum volume $V_*$ and a maximum density $\rho_c/2$, with finite curvature: the big bang singularity of the dust model is replaced by a material boundary.
\item The linear model reproduces the effective Friedmann equation of Loop Quantum Cosmology, with the critical density at the decoupling surface $\kappa=0$, but not its dynamics. No theory of the class \eqref{eq:E} bounces in flat FLRW with $\kappa>0$ (Theorem~\ref{prop:nobounce}), and for dust the difference from LQC is a single isotropic pressure term (Proposition~\ref{prop:lqc}). This term has a precise origin. Effective LQC is the conservative completion of the linear model: the same energy function with the physical rather than the dressed number conserved. The pressure $-\rho^2/\rho_c$ is the compression work of the medium, which $\kappa(R,T)$ does not return. The fold sits exactly at the density $\rho_c/2$ where the deformed constraint algebra of LQC changes signature. Bounces exist only in the repulsive phase $\kappa<0$: in the linear model they are exact solutions driven by $\Lambda$ against a negative effective energy, in a sector disconnected from the attractive branch.
\item In the reduced dust model, the symmetric quantization of the constraint is a first-order transport equation of Koopman--von Neumann type, independent of $\hbar$ and solved exactly; in the covariant vacuum theory the free propagator is purely mixed. The adjoint field is the response variable of the theory and suggests an interpretation as the difference field of an open-system action, consistent with the non-conservation of matter, with the polarizable medium as a natural candidate for the environment (Section~\ref{sec:quantum}).
\end{enumerate}
The comparison with Loop Quantum Cosmology deserves emphasis. A Friedmann equation of the form $H^2\propto\rho(1-\rho/\rho_c)$ is the hallmark of effective LQC, and in that framework, as in braneworld, mimetic and energy--momentum-squared models, it signals a bounce that replaces the big bang. The linear $\kappa(T)$ model reproduces this equation exactly, and places the critical density on the decoupling surface $\kappa=0$, where gravity switches off. It nevertheless does not bounce. The difference lies entirely in which particle number is conserved, and can be isolated as a single pressure term $-\rho^2/\rho_c$, the work of the medium (Proposition~\ref{prop:lqc} and Section~\ref{sec:lqccomp}). The quadratic Friedmann equation therefore does not by itself imply a bounce. In $\kappa(R,T)$ the high-density regime ends at a material fold instead, at the density where LQC changes signature, and a bounce is possible only in the repulsive phase.

The work is organized as follows. Section~\ref{sec:chen} presents Chen's action and its relation to adjoint and doubled actions. Section~\ref{sec:reduction} performs the canonical reduction. Section~\ref{sec:mediumsec} derives the non-conservation of matter and the polarizable-medium form of the theory (results 2 and 3). Section~\ref{sec:physics} treats the material fold and the comparison with Loop Quantum Cosmology (results 4 and 5). Section~\ref{sec:quantum} discusses quantization in the reduced model and the quadratic vacuum sector. Section~\ref{sec:conclusions} summarizes the results and outlines a programme to derive Chen's action from gravity coupled to a polarizable medium.

\paragraph{Conventions and scope.} We use signature $(+---)$, $\dd s^2=N(t)^2\dd t^2-e^{2q(t)}\dd\bm x^2$, $H=\dot q/N$, $V=e^{3q}$, and unit fiducial comoving volume. The matter source is a prescribed homogeneous perfect fluid $p=w\rho$ with constant $w$; it has no independent action. All cosmological statements refer to flat FLRW. Outside the canonical expressions, dots denote proper time ($N=1$).
\newpage

\section{Chen's extended action}\label{sec:chen}
This section presents Chen's construction \cite{Chen} and situates it in the literature on adjoint and doubled actions. The action and its recovery theorem are due to Chen; the subsequent sections consist of deductions from it that are not part of \cite{Chen}.

\subsection{The action and exact projected recovery}
For a specified smooth, local and covariant stress assignment $T_{\mu\nu}[g,\chi]$ and a smooth coupling $\kappa\in C^\infty(\mathbb R^2)$, Chen introduces a contravariant multiplier $a^{\mu\nu}$, with all sixteen components independent, and the action
\begin{equation}\label{eq:S}
\begin{split}
S_{\rm Ch}[g,\chi,a]&=\int\dd^4x\sqrt{|g|}\;a^{\mu\nu}\Bigl[R_{\mu\nu}-\tfrac12Rg_{\mu\nu}-\Lambda g_{\mu\nu}-\kappa\bigl(R,g^{\alpha\beta}T_{\alpha\beta}[g,\chi]\bigr)\,T_{\mu\nu}[g,\chi]\Bigr]\\
&=\int\dd^4x\sqrt{|g|}\;a^{\mu\nu}E_{\mu\nu} .
\end{split}
\end{equation}
It follows by integrating the prescribed derivative $\partial\mathcal L/\partial a^{\mu\nu}=\sqrt{|g|}\,E_{\mu\nu}$. No division by $\kappa$ and no inversion of the trace map occur, so the action is smooth at $\kappa=0$. For symmetric $T_{\mu\nu}$ the antisymmetric part of $a^{\mu\nu}$ multiplies a symmetric tensor and drops out. Writing $h_{\mu\nu}=\delta g_{\mu\nu}$, $\tau_{\mu\nu}=\delta T_{\mu\nu}$ and $b^{\mu\nu}=\delta a^{\mu\nu}$, the first variation is
\begin{equation}\label{eq:chenvar}
\delta S_{\rm Ch}=\int\dd^4x\sqrt{|g|}\Bigl[b^{\mu\nu}E_{\mu\nu}+a^{\mu\nu}\,\delta E_{\mu\nu}+\tfrac12 g^{\alpha\beta}h_{\alpha\beta}\,a^{\mu\nu}E_{\mu\nu}\Bigr],
\end{equation}
with
\begin{equation}
\delta E_{\mu\nu}=\delta R_{\mu\nu}-\tfrac12\delta R\,g_{\mu\nu}-\bigl(\tfrac12R+\Lambda\bigr)h_{\mu\nu}-(\kappa_R\delta R+\kappa_T\delta T)\,T_{\mu\nu}-\kappa\,\tau_{\mu\nu},
\end{equation}
where $\kappa_R,\kappa_T$ are the partial derivatives of the coupling and $\delta T=\delta g^{\alpha\beta}T_{\alpha\beta}+g^{\alpha\beta}\tau_{\alpha\beta}$ retains the full metric and matter dependence of the source. Stationarity under $b$ gives $E_{\mu\nu}=0$. Stationarity under $(g,\chi)$ gives the weak adjoint equations, which are linear and homogeneous in $a$. Chen's recovery theorem states
\begin{equation}\label{eq:recovery}
\bigl[\exists\,a:\ (g,\chi,a)\ \text{stationary}\bigr]\iff E_{\mu\nu}[g,\chi]=0,
\end{equation}
and every solution of \eqref{eq:E} lifts with $a=0$. The contracted Bianchi identity then yields $\nabla^\mu(\kappa T_{\mu\nu})=0$ in undivided form, which at $\kappa=0$ reduces to the compatibility condition $T_{\mu\nu}\nabla^\mu\kappa=0$ of \cite{Teruel2026}. The first variation, the projection theorem and the contracted Bianchi identity are formally verified in Lean~4 \cite{Chen}.

Chen is explicit about the scope of \eqref{eq:recovery}. It is a projection theorem, $\mathrm{proj}_{(g,\chi)}\mathrm{Crit}(S_{\rm Ch})=\mathrm{Sol}(E)$, not a bijection between solution spaces. It neither eliminates the adjoint sector nor establishes equality of physical degrees of freedom, provides a quantum effective action, or supplies microscopic matter equations beyond the specified stress assignment.

\subsection{Relation to adjoint and doubled actions}
Chen presents the construction as an instance of the formal-Lagrangian principle \cite{AthertonHomsy,Ibragimov,Peng}, within a reconstruction viewpoint developed in companion work \cite{ChenTutorial,ChenRecon}. For any system $E_a[q]=0$, the action $S[q,\lambda]=\int\lambda^aE_a[q]$ yields the original equations together with their adjoint equations. Ibragimov used this to define nonlinear self-adjointness and to extend Noether's theorem to non-variational equations \cite{Ibragimov}.

The same structure appears in physics under several names. Bateman's Lagrangian for the damped oscillator \cite{Bateman},
\begin{equation}\label{eq:bateman}
L_{\rm B}=m\dot x\dot y+\tfrac\gamma2(x\dot y-\dot xy)-kxy\;\simeq\;-\,y\,\bigl(m\ddot x+\gamma\dot x+kx\bigr),
\end{equation}
is, up to a total derivative, the adjoint variable $y$ times the damped equation. Its variation gives the damped oscillator for $x$ and the time-reversed, amplified oscillator for $y$. Chen's action is the field-theoretic analogue for $\kappa(R,T)$: $a^{\mu\nu}$ plays the role of $y$, and its equation is the adjoint of the linearized gravitational equations. In statistical physics the multiplier is the response field of the Martin--Siggia--Rose formalism \cite{MSR}. In the classical path integral of Gozzi and collaborators it is the variable conjugate to the classical trajectory, supplemented by ghosts for the Jacobian \cite{Gozzi}. In Galley's formulation of nonconservative mechanics \cite{Galley} the variables are doubled, $x_\pm$, and the action is $S[x_+]-S[x_-]+K[x_+,x_-]$, where $K$ carries the nonconservative forces; to first order in $x_-$ this is again an adjoint variable times the equations of motion. Chen's action does not split off a conservative part, even where one exists, as in the Einstein--Hilbert sector: the entire dynamics resides in the linear term. Section~\ref{sec:adjoint} shows that, for barotropic matter, an ordinary effective-fluid action exists where the constitutive reconstruction is regular, and that Chen's action remains available where it is not.

This last point distinguishes the construction from ordinary actions for related theories. The action $\int\sqrt{-g}\,[R/2\kappa_0+\phi(R,T)\mathcal L_m]$ of \cite{PerezPena} is variational in the usual sense, but its field equations contain additional terms. In the subclass $\phi(T)$ they read $G_{\mu\nu}=\kappa_0\phi(T)T_{\mu\nu}-2\kappa_0\phi_T(T_{\mu\nu}+\Theta_{\mu\nu})\mathcal L_m$.
Here $\Theta_{\mu\nu}:=g^{\alpha\beta}\delta T_{\alpha\beta}/\delta g^{\mu\nu}$. The additional term vanishes in particular cases, including constant $\phi$ and the stated dust prescription with on-shell $\mathcal L_m=p=0$. The latter requires a specified variational fluid formulation; setting the matter action to zero before variation would not reproduce it. This comparison shows that the displayed nonminimal-coupling ansatz does not generically yield the target equations. It does not establish a no-go theorem for all ordinary actions with additional variables or constraints. Because its existence follows from a general principle, its value for $\kappa(R,T)$ must be shown by what it does: whether its canonical structure is consistent and what it reveals (Sections~\ref{sec:reduction}--\ref{sec:physics}), what quantum theory it defines in the reduced and quadratic sectors (Section~\ref{sec:quantum}), and whether the doubled form is merely available or natural for this theory (Sections~\ref{sec:medium} and~\ref{sec:adjoint}).

\section{Canonical reduction of Chen's action}\label{sec:reduction}
The first test of an action is whether it defines a consistent Hamiltonian system. We restrict Chen's action to flat FLRW with a prescribed barotropic fluid, keeping the coupling $\kappa(T)$ arbitrary, and carry out the Dirac analysis. The multiplier components become ordinary canonical variables, and the constraint algebra splits into a first-class pair, associated with time reparametrization, and a set of second-class constraints that tie the multiplier to the matter density. The outcome is a reduced Hamiltonian with a remarkable property: on every regular branch it no longer contains the coupling. This identifies $F'(\rho)=0$ as the locus where the matter reconstruction and the second-class elimination used here can cease to be regular.

\subsection{Minisuperspace Lagrangian}
For $\kappa=\kappa(T)$ and $T=(1-3w)\rho=:s\rho$ define the effective source
\begin{equation}\label{eq:F}
F(\rho):=\kappa(s\rho)\,\rho .
\end{equation}
The independent homogeneous components of \eqref{eq:E} are
\begin{align}
E_0&=3H^2-\Lambda-F(\rho),\label{eq:E0}\\
E_s&=\frac{2\dot H}{N}+3H^2-\Lambda+wF(\rho),\label{eq:Es}
\end{align}
since $G_{00}/N^2=3H^2$, $G_{ij}/g_{ij}=2\dot H/N+3H^2$, $T_{00}/N^2=\rho$ and $T_{ij}/g_{ij}=-p$. With homogeneous multiplier components $A$ and $C$ normalized so that
\begin{equation}
S_{\rm mini}=\int\dd t\,NV\bigl(AE_0+CE_s\bigr),
\end{equation}
and using $2VC\dot H=\frac{\dd}{\dd t}(2VCH)-2VH\dot C-2\dot VCH$ with $\dot V=3NVH$, we obtain up to a total derivative
\begin{equation}\label{eq:L}
L=V\left[\frac{3(A-C)\dot q^2-2\dot q\dot C}{N}-N\bigl((A+C)\Lambda+(A-wC)F(\rho)\bigr)\right].
\end{equation}

\subsection{Momenta, constraints and Hamiltonian}
The momenta are
\begin{equation}\label{eq:mom}
p_q=\frac VN\bigl[6(A-C)\dot q-2\dot C\bigr],\qquad
p_C=-\frac{2V}{N}\dot q,\qquad p_A=p_\rho=p_N=0,
\end{equation}
invertible in the velocities $(\dot q,\dot C)$ for $V>0$ and non-degenerate lapse $N\neq0$ The Legendre transform gives $H_{\rm can}=N\cH$ with
\begin{equation}\label{eq:Ham}
\cH=-\frac{p_qp_C}{2V}-\frac{3(A-C)p_C^2}{4V}+V\bigl[(A+C)\Lambda+(A-wC)F(\rho)\bigr].
\end{equation}
Preservation of the primary constraints $p_N,p_A,p_\rho$ gives
\begin{align}
\cH&\approx0,\\
X:=\frac{\partial\cH}{\partial A}&=-\frac{3p_C^2}{4V}+V(\Lambda+F)\approx0,\label{eq:X}\\
Y:=\frac{\partial\cH}{\partial\rho}&=V(A-wC)F'(\rho)\approx0.\label{eq:Y}
\end{align}
With $p_C=-2VH$, the constraint $X$ is $-VE_0$.

\subsection{Regular branches and canonical universality}
On a branch with $F'(\rho)\neq0$, $Y=0$ gives $A=wC$ and $X=0$ fixes
\begin{equation}\label{eq:J}
F(\rho)=J(q,p_C):=\frac{3p_C^2}{4V^2}-\Lambda .
\end{equation}
The constraints $(p_A,\,A-wC,\,p_\rho,\,X)$ are second class, since $\{p_A,A-wC\}=-1$ and $\{p_\rho,X\}=-VF'\neq0$.  Their Poisson-bracket matrix has determinant $V^2[F'(\rho)]^2$, which is nonzero on the branch considered. Eliminating them leaves $\Theta_{\rm red}=p_q\dd q+p_C\dd C$ and
\begin{equation}\label{eq:Hred}
\cH_{\rm red}=-\frac{p_qp_C}{2V}+C\left[\frac{3(1-w)p_C^2}{4V}+(1+w)\Lambda V\right].
\end{equation}
Eliminating the four second-class constraints reduces the ten-dimensional phase space to six dimensions, including $(N,p_N)$. On the regular part of the remaining constraint surface, $(p_N,\mathcal H_{\rm red})$ is a first-class pair and removes four further dimensions, leaving one physical canonical pair. Before second-class reduction, the corresponding first-class generator requires a suitable extension of $\mathcal H$ by terms proportional to the second-class constraints.
\begin{theorem}[Canonical universality]\label{thm:universal}
For a prescribed homogeneous perfect fluid with $p=w\rho$ and constant $w$, on any branch with $F'(\rho)\neq0$ the reduced constraint \eqref{eq:Hred} is independent of the functional form of $\kappa(T)$. The coupling enters only through the reconstruction $F(\rho)=J$, its range and the choice of inverse branch.
\end{theorem}
\begin{proof}
The only explicit occurrence of $F$ in \eqref{eq:Ham} is $V(A-wC)F$, which vanishes identically once $Y=0$ has fixed $A=wC$. The elimination of the remaining second-class pair does not reintroduce $F$.
\end{proof}

Geometrically, on each regular branch the cosmological dynamics of \eqref{eq:E} obeys the equations of general relativity with an effective fluid of density $F$ and pressure $wF$. The function $\kappa$ determines which matter density realises a given geometry, not the geometry itself. This conclusion holds for the field equations independently of the action. What the action adds is its canonical expression: the coupling disappears from the reduced Hamiltonian and survives only in a second-class constraint. The reduction therefore identifies $F'(\rho)=0$ as a possible degeneracy of the local matter reconstruction. It does not exclude singularities associated with other boundaries or sectors. Section~\ref{sec:fold} analyses the non-degenerate fold of the linear dust model and establishes its obstruction to regular Hamiltonian continuation.

This is an identification of the geometric sector, not an equivalence of theories. The effective fluid is tied to the physical matter by the map $\rho\mapsto F(\rho)$, and two features of that map prevent a global identification with general relativity. First, on critical surfaces $\kappa=0$ the effective source vanishes while the physical matter does not; such a configuration is not vacuum general relativity with the same matter. These surfaces obstruct a global Einstein-frame reformulation of the theory \cite{Teruel2026}. Since $F'=T\kappa'(T)\neq0$ there in general, the canonical reduction remains regular at $\kappa=0$, and the obstruction concerns the physical identification, not the constraint structure. Second, at a fold $F'=0$ the map is not invertible, and no physical matter realises effective densities beyond its extremum (Section~\ref{sec:fold}). The fold is therefore a second obstruction of the same kind, and unlike the critical surface it is also visible in the constraint structure of Chen's action. The effective-source rewriting remains valid wherever the original equations are defined. Its inversion requires distinguishing the scalar FLRW reconstruction, which can remain regular at $\kappa=0$, from the full tensorial matter map, which degenerates there. Non-degenerate folds instead obstruct the local scalar reconstruction and the canonical reduction used here.

The comparison with Rastall's theory \cite{Rastall} makes the point sharper. There $\nabla_\mu T^{\mu\nu}=\lambda\nabla^\nu R$, the trace of the field equations fixes $R\propto T$, and the theory is general relativity with the effective source $T_{\mu\nu}-\frac{\kappa\lambda}{4\kappa\lambda-1}\,T g_{\mu\nu}$. Since this map is linear, with constant coefficients, and invertible for every admissible value of the parameter, the identification holds globally, and Rastall gravity is general relativity with a redefined fluid \cite{Visser2018}; what remains open there is which of the two tensors describes the physical matter \cite{Darabi2018}. In $\kappa(R,T)$ the map $T_{\mu\nu}\mapsto\kappa(R,T)T_{\mu\nu}$ is nonlinear and depends on the state. It fails to be invertible on critical surfaces and at folds, and these loci carry the content of the theory beyond the effective-fluid identification.

The regularity criterion is $F'(\rho)=\kappa(T)+T\kappa'(T)\neq0$, not $\kappa\neq0$. The coupling can vanish at a regular point, and $F'$ can vanish at non-zero coupling.

\section{\texorpdfstring{$\kappa(R,T)$}{kappa(R,T)} as gravity in a polarizable medium}\label{sec:mediumsec}
Theorem~\ref{thm:universal} separates two roles that are mixed in the field equations: on each regular branch the geometry obeys the equations of general relativity with an effective fluid, while the coupling governs only the relation between that fluid and the physical matter. This section and the next follow the consequences of that separation. The Bianchi invariant of the reduced system fixes how matter exchanges energy with the geometry (Section~\ref{sec:nonconservative}), an exchange that can be rewritten exactly as gravity in a polarizable medium (Section~\ref{sec:medium}). The reformulation holds at the level of the field equations, and its language is used in the rest of the paper.

\subsection{Non-conservative matter}\label{sec:nonconservative}
The Bianchi identity applied to \eqref{eq:E} gives $\nabla^\mu(\kappa T_{\mu\nu})=0$. In FLRW this reads
\begin{equation}\label{eq:bianchi}
\dot F+3H(1+w)F=0,\qquad FV^{1+w}=M_w=\text{const},
\end{equation}
which is the statement $XM=0$ of Section~\ref{sec:quantum} for dust. Equivalently, for the matter density itself,
\begin{equation}\label{eq:Gamma}
\nabla^\mu T_{\mu\nu}=-T_{\mu\nu}\nabla^\mu\ln|\kappa|,\qquad
\dot\rho+3H(1+w)\rho=\Gamma\rho,\quad \Gamma=-\frac{\dot\kappa}{\kappa}.
\end{equation}
The divided expressions hold where $\kappa\neq0$.
At a zero of the coupling, the appropriate balance equation is
the undivided identity $\nabla^\mu(\kappa T_{\mu\nu})=0$.

\begin{proposition}[Incompatibility with ordinary conservation]\label{prop:conservation}
Let $\kappa=\kappa(T)$, constant $w$, $\rho\neq0$, in comoving flat FLRW. If both $\nabla^\mu(\kappa T_{\mu\nu})=0$ and $\nabla^\mu T_{\mu\nu}=0$ hold, then $(1+w)H\rho=0$ unless $\kappa'(T)=0$ or $w=1/3$.
\end{proposition}
\begin{proof}
Both laws together give $T_{\mu\nu}\nabla^\mu\kappa=0$. For a homogeneous comoving fluid $\nabla_\nu\kappa=u_\nu\dot\kappa$, so $\rho\,\dot\kappa\,u_\nu=0$ and $\dot\kappa=0$. Since $\dot\kappa=\kappa'(T)(1-3w)\dot\rho$, this forces $\dot\rho=0$ when $\kappa'\neq0$ and $w\neq1/3$, and ordinary conservation then gives $3H(1+w)\rho=0$.
\end{proof}
Thus, under the hypotheses of Proposition~\ref{prop:conservation},
ordinary conservation is excluded on evolving intervals where
$\kappa_T\neq0$ and $w\neq-1,1/3$. The theory fixes the rate at which it exchanges energy with the geometry. As single components, radiation ($T=0$) and vacuum energy ($w=-1$) are unaffected; in mixtures $\kappa$ depends on the total trace, and how the exchange is shared among components is not fixed by the field equations. The factor $1/\kappa$ in $\Gamma$ belongs to the rewritten form \eqref{eq:Gamma}; the conserved quantity $FV^{1+w}$ of \eqref{eq:bianchi} is regular at $\kappa=0$, in agreement with \cite{Teruel2026}. This is the $\kappa(R,T)$ counterpart of the non-conservation familiar from $f(R,T)$ gravity \cite{Harko}, here in closed form. A conserved number nevertheless exists. For an isentropic barotropic fluid satisfying $d\rho=(\rho+p)\,dn/n$, the energy projection of $\nabla^\mu(\kappa T_{\mu\nu})=0$ is equivalent to
$\nabla_\mu(Nu^\mu)=0$ for the dressed density

\begin{equation}\label{eq:dressed}
N=nD(n),\qquad
\ln|D(n)|=\int\frac{\rho}{\rho+p}\,d\ln|\kappa|,
\end{equation}
which follows from $d\ln N=d(\kappa\rho)/[\kappa(\rho+p)]$ and the first law at constant specific entropy, $d\rho=(\rho+p)\,dn/n$.
This construction does not replace the transverse Euler equations.
For curvature-dependent couplings, the flow-line integral need not
define a local universal constitutive function $\varepsilon(N)$. For $\kappa(T)$, $D$ is a function of $n$; for $\kappa(R,T)$ the integral is taken along the flow lines. For dust $N=(\kappa/\kappa_0)\,n$, and $FV=$ const is the statement that $N$ is conserved. The geometry and the flow are those of general relativity with a Lagrangian fluid whose Noether charge is $N$; matter is non-conservative only in the sense that the physical number is not this charge. The physical number then evolves as $\dot n/n+3H=\Gamma/(1+w)$, so $\Gamma$ is the fractional rate of change of the number only for dust. Where $\kappa<0$ the dressed number of dust is negative: $N$ is a signed conserved density, and its reading as a count of free plus bound particles applies where $\kappa>0$.

Equation \eqref{eq:Gamma} has the form of the balance equations of cosmologies with matter creation \cite{Prigogine}, but the physics differs. The field equations, and with them the reduced dynamics \eqref{eq:geom}, are invariant under time reversal, while $\Gamma=-\dot\kappa/\kappa$ is odd. On the low-density branch of the linear model with $\lambda>0$, matter is destroyed in expansion ($\Gamma<0$) and created in contraction ($\Gamma>0$); on the high-density branch the signs are reversed. The balance law is therefore time-reversal symmetric and imposes no sign-definite rate, so the second-law condition $\Gamma\geq0$ of irreversible particle creation does not apply to it; no entropy-production law for the medium is implied. Section~\ref{sec:medium} gives the natural setting for an exchange of this kind.

\subsection{Induced source and gravitational susceptibility}\label{sec:medium}
The non-conservation of matter admits an exact algebraic reformulation. For a reference coupling $\kappa_0>0$, define
\begin{equation}\label{eq:Tenv}
T^{\rm env}_{\mu\nu}:=\chi_g T_{\mu\nu},\qquad
\chi_g(R,T):=\frac{\kappa(R,T)}{\kappa_0}-1.
\end{equation}
The field equations and their Bianchi identity become
\begin{equation}\label{eq:medium}
G_{\mu\nu}-\Lambda g_{\mu\nu}=\kappa_0\bigl(T_{\mu\nu}+T^{\rm env}_{\mu\nu}\bigr),\qquad
\nabla^\mu\bigl(T_{\mu\nu}+T^{\rm env}_{\mu\nu}\bigr)=0.
\end{equation}
The second component is an \emph{induced effective source}, not an independently established material environment. Its divergence compensates that of ordinary matter, but this identity alone does not establish microscopic energy transfer, dissipation, or entropy production. In the constitutive interpretation, $\chi_g$ plays the role of a nonlinear gravitational susceptibility, $T^{\rm env}_{\mu\nu}$ that of an induced polarization-like contribution, and $\Gamma$ parametrizes the effective exchange between the two algebraically defined sectors.

The singular loci of the theory acquire a direct meaning in this language:
\begin{itemize}
\item \emph{Critical surfaces} $\kappa=0$ are surfaces of total screening, $T^{\rm env}=-T$: the induced contribution cancels the matter source. This is the gravitational screening identified in \cite{Teruel2026}.
\item \emph{The repulsive phase} $\kappa<0$ is over-screening: the induced stress exceeds the matter stress and reverses the sign of the total source, formally reminiscent of media with negative permittivity.
\item \emph{The fold} is a saturation-like turning point: the total scalar source $F=\kappa\rho$ that matter and the induced contribution supply together is bounded above by $F_*$ in the linear dust model. Beyond $\rho_*$, increasing $\rho$ decreases the total source. This is an analogy with nonlinear dielectric saturation, not an identification with a microscopic saturation mechanism.
\end{itemize}
The distinction between saturation at $F'(\rho_*)=0$ and total screening at $\kappa(\rho_c)=0$ is illustrated in Fig.~\ref{fig:critical-surfaces}. The reformulation is exact and adds no assumptions; it is a change of description, not a model of the medium. Its value is that it suggests a constitutive language for the open-system interpretation of Section~\ref{sec:adjoint}, while leaving the existence of an independent environment unproven.

\subsection{Structural analogy with Maxwell theory in matter}\label{sec:maxwell}
For a homogeneous, isotropic linear dielectric, Gauss's law and the electric constitutive relation are
\begin{equation}\label{eq:maxwell_dielectric}
\boldsymbol\nabla\!\cdot\mathbf D=\rho_{\rm free},\qquad
\mathbf D=\varepsilon_0(1+\chi_e)\mathbf E,
\end{equation}
so that
\begin{equation}\label{eq:maxwell_susceptibility}
\boxed{\boldsymbol\nabla\!\cdot[\varepsilon_0(1+\chi_e)\mathbf E]=\rho_{\rm free}.}
\end{equation}
The corresponding gravitational field equation can be displayed as
\begin{equation}\label{eq:gravity_susceptibility}
\boxed{G_{\mu\nu}-\Lambda g_{\mu\nu}=\kappa_0[1+\chi_g(R,T)]T_{\mu\nu}.}
\end{equation}
These equations exhibit a susceptibility-dependent response, although the dielectric susceptibility multiplies a field inside a differential operator whereas $\chi_g$ multiplies the gravitational source. A closer source-level analogy follows from $\mathbf P=\varepsilon_0\chi_e\mathbf E$ and $\rho_{\rm bound}=-\boldsymbol\nabla\cdot\mathbf P$:
\begin{equation}\label{eq:maxwell_bound}
\varepsilon_0\boldsymbol\nabla\cdot\mathbf E=\rho_{\rm free}+\rho_{\rm bound},
\qquad
G_{\mu\nu}-\Lambda g_{\mu\nu}=\kappa_0(T_{\mu\nu}+T^{\rm env}_{\mu\nu}).
\end{equation}
In covariant electromagnetism the induced current may be written $J^\nu_{\rm bound}=\nabla_\mu\mathcal M^{\mu\nu}$ with antisymmetric $\mathcal M^{\mu\nu}$. Unlike this current, $T^{\rm env}_{\mu\nu}=\chi_g T_{\mu\nu}$ is generally not separately conserved. Table~\ref{tab:maxwell_gravity} makes both the analogy and its limits explicit. The analogy is exact at the level of the conserved number. For dust, Eq.~\eqref{eq:dressed} reads
\begin{equation}\label{eq:boundnumber}
\nabla_\mu\bigl(n\,u^\mu+\chi_g\,n\,u^\mu\bigr)=0 ,
\end{equation}
the counterpart of $\nabla_\nu(J^\nu_{\rm free}+J^\nu_{\rm bound})=0$: the induced sector carries the part $\chi_g n$ of the conserved number, and only the total is conserved.

This gives content to the motivation of the original formulation, which presented $\kappa(R,T)$ as a gravitational analogue of Maxwell's displacement current \cite{Perez}. In both theories the field equations are completed by a source term required by a differential identity on the field side. In electrodynamics, the divergence of the Ampere--Maxwell equation, together with Gauss's law, yields charge continuity. In gravity, the Bianchi identity $\nabla^\mu G_{\mu\nu}=0$ forces the total source to be conserved, and $T^{\rm env}_{\mu\nu}$ restores this when matter is non-conservative. The correspondence is with the material part of the displacement current, the polarization current $\partial_t\mathbf P$, not with its vacuum part $\varepsilon_0\partial_t\mathbf E$. This algebraic rewriting introduces no independent fields and leaves the degree-of-freedom content of the original equations unchanged. One difference remains: in a dielectric, free and bound charge are conserved separately, whereas here only their sum is, and the two exchange number at the rate $\Gamma$. The closer electromagnetic picture is therefore a medium in which charges are captured and released, rather than an ideal dielectric.

\begin{table}[htbp]
\centering\small
\caption{Constitutive analogy between Maxwell theory in matter and $\kappa(R,T)$ gravity. The correspondence is formal, not a microscopic equivalence.}\label{tab:maxwell_gravity}
\begin{tabular}{@{}p{0.22\textwidth}p{0.33\textwidth}p{0.35\textwidth}@{}}
\toprule
Structure & Maxwell in matter & $\kappa(R,T)$ gravity\\
\midrule
Field equation & $\nabla_\mu H^{\mu\nu}=J^\nu_{\rm free}$ & $G_{\mu\nu}-\Lambda g_{\mu\nu}=\kappa T_{\mu\nu}$\\[3pt]
Susceptibility & $\mathbf D=\varepsilon_0(1+\chi_e)\mathbf E$ & $\kappa=\kappa_0(1+\chi_g)$\\[3pt]
Induced source & $J^\nu_{\rm bound}=\nabla_\mu\mathcal M^{\mu\nu}$ & $T^{\rm env}_{\mu\nu}=\chi_g T_{\mu\nu}$\\[3pt]
Total source & $J_{\rm free}+J_{\rm bound}$ & $T_{\mu\nu}+T^{\rm env}_{\mu\nu}$\\[3pt]
Differential response & $\partial D_i/\partial E_j$ & $\partial(\kappa T_{\mu\nu})/\partial T_{\alpha\beta}$\\[3pt]
Conservation & $\nabla_\nu J^\nu_{\rm free}=0$ & $\nabla^\mu(\kappa T_{\mu\nu})=0$\\[3pt]
Conserved current & $J_{\rm free}+J_{\rm bound}$ & $(1+\chi_g)\,n\,u^\mu$ (dust)\\[3pt]
Microscopic medium & Polarizable charges and dipoles & Not specified\\
\bottomrule
\end{tabular}
\end{table}

\subsection{Tensorial differential susceptibility and the fold}\label{sec:tensorial}
For $\kappa=\kappa(T)$, varying the stress tensor at fixed metric gives
\begin{equation}\label{eq:constitutive_variation}
\delta(\kappa T_{\mu\nu})=\mathcal K_{\mu\nu}{}^{\alpha\beta}\delta T_{\alpha\beta},\qquad
\mathcal K_{\mu\nu}{}^{\alpha\beta}=\kappa\,\delta_{(\mu}^{\alpha}\delta_{\nu)}^{\beta}+\kappa_T T_{\mu\nu}g^{\alpha\beta}.
\end{equation}
On the ten-dimensional space of symmetric tensors in four dimensions this rank-one update has characteristic polynomial
\begin{equation}\label{eq:constitutive_det}
\det(zI-\mathcal K)=(z-\kappa)^9[z-(\kappa+T\kappa_T)],\qquad
\boxed{\det\mathcal K=\kappa^9(\kappa+T\kappa_T).}
\end{equation}
For $T\neq0$ the nine-dimensional traceless subspace has eigenvalue $\kappa$, while the direction spanned by $T_{\mu\nu}$ has eigenvalue $\kappa+T\kappa_T$. At $T=0$ these eigenvalues coincide and diagonalizability need not follow, but the characteristic polynomial remains valid.

For dust with $T=\rho$ and $\kappa(T)=\kappa_0-\lambda T$, the second characteristic value is
\begin{equation}\label{eq:constitutive_fold}
\kappa+T\kappa_T=\kappa_0-2\lambda\rho
=\frac{\dd}{\dd\rho}[\kappa(\rho)\rho]=F'(\rho).
\end{equation}
Thus the material fold at $\rho_*=\kappa_0/(2\lambda)$ is a degeneracy of the differential constitutive map even though $\kappa(\rho_*)=\kappa_0/2>0$. In the homogeneous Dirac reduction below, the same factor occurs in $\{p_\rho,X\}=-VF'(\rho)$. This is a precise connection between the constitutive degeneracy and one entry of the reduced constraint matrix; it is not yet a covariant theorem about the rank of the full constraint system.

For general $\kappa(R,T)$ the variation also contains $\kappa_R T_{\mu\nu}\delta R$, so $\mathcal K$ is only the partial matter response at fixed curvature. Finally, Chen's adjoint $a^{\mu\nu}$ is a response variable enforcing the equations, not the induced source $T^{\rm env}_{\mu\nu}$. The exact reformulation does not itself derive an environment or a Schwinger--Keldysh influence functional.

\section{Material folds and Loop Quantum Cosmology}\label{sec:physics}
The degeneration of the second-class constraint identifies where the matter description fails (Section~\ref{sec:fold}). The universality of the geometric sector constrains which cosmologies the whole $\kappa(R,T)$ class can produce, which we use to compare it with Loop Quantum Cosmology (Section~\ref{sec:lqccomp}).

\subsection{The material fold}\label{sec:fold}
\begin{figure}[htbp]
\centering
\includegraphics[width=0.96\linewidth]{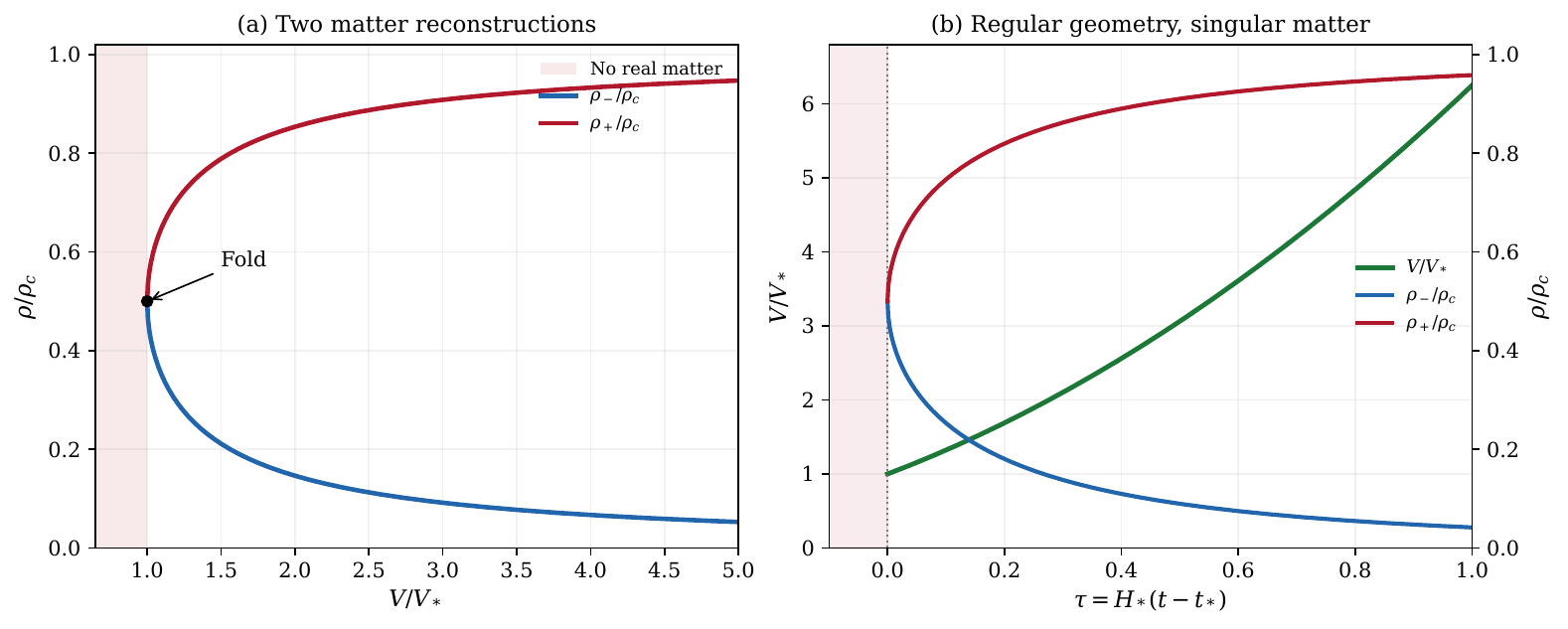}
\caption{Material fold in the linear dust model: the two inverse density branches and their proper-time evolution. The geometric evolution remains regular at the fold while the reconstructed matter derivative diverges.}
\label{fig:material-fold}
\end{figure}

Take dust, $\kappa_0>0$, $\lambda>0$, $M:=M_0>0$ and
\begin{equation}
F(\rho)=\kappa_0\rho-\lambda\rho^2,\qquad F_*=\frac{\kappa_0^2}{4\lambda}\ \text{at}\ \rho_*=\frac{\kappa_0}{2\lambda}.
\end{equation}
Since $F=M/V$,
\begin{equation}\label{eq:branches}
V\ge V_*:=\frac{4\lambda M}{\kappa_0^2},\qquad
\rho_\pm(V)=\frac{\kappa_0}{2\lambda}\left(1\pm\sqrt{1-\frac{V_*}{V}}\right),
\end{equation}
two material reconstructions of the \emph{same} geometry. The density obeys
\begin{equation}\label{eq:rhodot}
\dot\rho=-3H\rho\,\frac{\kappa_0-\lambda\rho}{\kappa_0-2\lambda\rho}.
\end{equation}
Faster-than-$a^{-3}$ dilution on the low-density branch and a divergence at $\rho_*$ are both manifest. Figure~\ref{fig:material-fold} displays the two inverse branches and their proper-time evolution.

\begin{proposition}[Fold]\label{prop:fold}
In the linear dust model:
\begin{enumerate}
\item The geometry is that of flat GR with conserved effective dust mass $M$,
\begin{equation}\label{eq:geom}
3H^2=\Lambda+\frac MV,\qquad \dot H=-\frac{M}{2V},\qquad \ddot V=3\Lambda V+\frac{3M}{2},
\end{equation}
and is analytic in proper time through $V=V_*$, with finite curvature and all derivatives finite.
\item If $H_*\neq0$, then $\rho_\pm-\rho_*=\pm\frac{\rho_c}{2}\sqrt{3|H_*|\,|t-t_*|}+O(|t-t_*|^{3/2})$ with $\rho_c=\kappa_0/\lambda$, so $\dot\rho$ and $\dot\kappa$, and hence $\Gamma$, diverge as $|t-t_*|^{-1/2}$.
\item For $\Lambda\ge0$, an expanding solution has no real-matter continuation to earlier times through the fold, continuous or not; a contracting one has none to later times.
\item Towards increasing $V$ both branches $\rho_\pm$ are admissible, with identical geometry: the reconstruction of the matter from the fold is two-valued, to the future for an expanding solution and to the past for a contracting one. Regular data with $\rho\neq\rho_*$ select one branch.
\end{enumerate}
\end{proposition}
\begin{proof}
(1) follows from Theorem~\ref{thm:universal} and \eqref{eq:bianchi}; the equation for $\ddot V$ has analytic coefficients. (2) follows by expanding \eqref{eq:branches} with $V-V_*=3H_*V_*(t-t_*)+O((t-t_*)^2)$. (3) For $\Lambda\geq0$, $H_*^2=(\Lambda+F_*)/3>0$. The unique geometric continuation has $V<V_*$, hence $F=M/V>F_*$, which no real $\rho$ attains. Equivalently, a differentiable crossing would require $F'(\rho_*)\dot\rho=-3H_*F_*=0$, which is impossible. (4) follows from \eqref{eq:branches}.
\end{proof}

The fold is thus a finite-curvature singularity of the matter sector. It is neither a curvature singularity nor a sudden singularity in the usual sense, since no derivative of the geometry diverges, and it is not a bounce. Expanding universes of the linear model on the low-density branch begin at finite volume, curvature and density $\rho_*=\rho_c/2$, with an infinite rate of energy exchange. For $\Lambda<0$ a turning point $H=0$ has $\dot H=-F/2<0$ and is a maximum of $V$. For $|\Lambda|<F_*$ the admissible interval is $V_*\le V\le M/|\Lambda|$; the upper end is a turning point of the volume, and the trajectory begins and ends at folds.

Homogeneous cosmology cannot reach the critical surface either. Since $M=FV$ is conserved, $F=\kappa\rho$ cannot vanish along a trajectory with $M\neq0$. The surface $\kappa=0$ therefore separates the sectors $M>0$ (attractive, $\kappa>0$) and $M<0$ (repulsive, $\kappa<0$), and no FLRW dust trajectory with $M\neq0$ reaches it; on the low-density branch the fold is reached first in any case. The sector $M=0$ contains, besides vacuum, the configuration $\rho\equiv\rho_c$ that lies on the critical surface at all times, with the geometry of vacuum GR with $\Lambda$; it is critical everywhere rather than bounded by a critical surface. It should not be confused with the fold: $F'(\rho_c)=-\kappa_0\neq0$, so the canonical reduction is regular there. Critical hypersurfaces that separate regions with $\kappa>0$ and $\kappa<0$, of the type studied in \cite{Teruel2026}, therefore require inhomogeneous configurations, such as stellar interiors, where $\rho$ varies in space. The two obstructions to a global effective-fluid identification thus belong to different regimes: the fold to cosmology, the critical surface to inhomogeneous matter distributions.

In the canonical description the fold is exactly where $\{p_\rho,X\}=-VF'$ vanishes: $Y$ no longer fixes $A=wC$ and the second-class structure degenerates. With the total Hamiltonian $H_T=N\cH+u_Ap_A+u_\rho p_\rho+u_Np_N$, a direct computation gives
\begin{equation}\label{eq:XH}
\{X,\cH\}=-\tfrac32(1+w)\,p_C\,F,\qquad \dot X=N\{X,\cH\}+u_\rho VF'(\rho).
\end{equation}
On a regular branch, $\dot X\approx0$ fixes $u_\rho=\dot\rho=-3NH(1+w)F/F'$, which is \eqref{eq:rhodot}.

Because $Y=V(A-wC)F'(\rho)$ factorizes, the constraint surface $\Sigma$ defined by $p_A=p_\rho=p_N=0$, $\cH=X=Y=0$ is the union of two components,
\begin{equation}\label{eq:components}
\Sigma_{\rm reg}:\ A-wC=0,\qquad \Sigma_{\rm fold}:\ \varphi:=F'(\rho)=0\ \ (A\ \text{free}),
\end{equation}
which intersect on $\{A=wC,\ F'=0\}$. On $\Sigma_{\rm fold}$ the constraint $Y$ is replaced by $\varphi$, which implies it, and the Dirac algorithm must preserve $\varphi$ itself.

\begin{proposition}[Dirac analysis at the fold]\label{prop:dirac}
Let $w$ be constant, $w\neq-1$, and let $\rho_*$ satisfy
$F'(\rho_*)=0$, $F_*:=F(\rho_*)\neq0$ and
$F''(\rho_*)<0$. Assume $V>0$ and a non-degenerate lapse.
\begin{enumerate}
\item On $\Sigma_{\rm fold}$, preservation of $\varphi$
requires $u_\rho=0$. Preservation of $X$ then imposes
the tertiary constraint $p_CF\approx0$, i.e.\ $H=0$.
For $\Lambda\neq-F_*$ this is incompatible with
$X\approx0$, so the component carries no consistent
evolution. For $\Lambda=-F_*$, $\cH\approx0$ also fixes
$C=0$, and the tertiary constraint is not preserved,
since
\[
\dot p_C=-N\,\frac{\partial\cH}{\partial C}
=NV(1+w)F_*\neq0.
\]
\item On $\Sigma_{\rm reg}$, at the intersection with
$F'=0$, preservation of $X$ with finite $u_\rho$
requires $p_C=0$, hence $H_*=0$ and $\Lambda=-F_*$.
Even in this exceptional case, no crossing with
continuously differentiable matter density exists,
as shown below.
\end{enumerate}
Hence no classical Hamiltonian trajectory of the
extended system with continuously differentiable
matter density and finite multipliers can remain
on the fold or pass through it while satisfying
all Dirac consistency conditions.
\end{proposition}

\begin{proof}
(1) On a trajectory confined to $\Sigma_{\rm fold}$,
\[
\dot\varphi=\{\varphi,H_T\}=F''(\rho)\,u_\rho,
\]
so preservation of $\varphi$ gives $u_\rho=0$.
Since $F'=0$, the $u_\rho$ term in \eqref{eq:XH}
vanishes, and
\[
\dot X=-\frac32N(1+w)p_CF.
\]
Preservation of $X$ therefore requires $p_C=0$,
and $X=V(\Lambda+F_*)$ requires $\Lambda=-F_*$.
For this value,
\[
\left.\cH\right|_{p_C=0,\,\rho=\rho_*}
=-V(1+w)CF_*,
\]
so $\cH\approx0$ fixes $C=0$. Moreover,
\[
\left.\frac{\partial\cH}{\partial C}
\right|_{p_C=0,\,\rho=\rho_*}
=-V(1+w)F_*,
\]
and hence
\[
\dot p_C=NV(1+w)F_*\neq0.
\]
Thus the condition $p_C=0$ cannot be preserved.
Preservation of $p_A$, $p_N$ and $\cH$ adds no
condition at this stage:
\[
\dot\cH=u_AX+u_\rho Y\approx0.
\]
The freedom in $u_A$ cannot remove the obstruction.

(2) On $\Sigma_{\rm reg}$,
\[
\dot X=-\frac32N(1+w)p_CF+u_\rho VF'.
\]
At $F'=0$, finite $u_\rho$ therefore requires
$p_C=0$, hence $H_*=0$ and $\Lambda=-F_*$.
To test a crossing, use proper time. Along any
putative classical continuation, the geometric
equations imply
\[
\dot F=-3(1+w)HF,\qquad
\dot H=-\frac12(1+w)F.
\]
Consequently,
\[
\dot F_*=0,\qquad
\ddot F_*=\frac32(1+w)^2F_*^2>0.
\]
However, continuous differentiability of $\rho$
and Taylor expansion at $\rho_*$ give
\[
F(\rho(t))-F_*=
\frac12F''(\rho_*)\dot\rho_*^2(t-t_*)^2
+o((t-t_*)^2).
\]
Since $F''(\rho_*)<0$, this contradicts the
strictly positive quadratic coefficient imposed
by the geometric equations. Thus no such
regular crossing exists.
\end{proof}

Proposition~\ref{prop:dirac} does not exclude trajectories of $\Sigma_{\rm reg}$ that approach the fold. They do so with $u_\rho=\dot\rho\to\infty$, and the limit point does not provide regular initial data from which the evolution could continue. The two statements are complementary: the fold is an endpoint of regular evolution and not a stratum that supports evolution. The calculation refers to the flat-FLRW sector with a prescribed fluid; it does not establish the behaviour of the covariant theory on general surfaces $F'=0$.

\begin{figure}[htbp]
\centering
\begin{tikzpicture}
\begin{axis}[
 name=source,width=7.1cm,height=6.0cm,
 xmin=0,xmax=1.25,ymin=-0.32,ymax=0.36,
 axis lines=left,axis line style={-Latex},
 xlabel={$x=\lambda\rho/\kappa_0$},ylabel={$\lambda F/\kappa_0^2$},
 xtick={0,0.5,1},xticklabels={$0$,$\frac12$,$1$},
 ytick={0,0.25},yticklabels={$0$,$\frac14$},
 title={(a) Effective source $F(\rho)$},title style={font=\small},
 tick label style={font=\scriptsize},clip=false]
 \addplot[blue,very thick,domain=0:1.22,samples=120]{x-x^2};
 \addplot[only marks,mark=*,red] coordinates {(0.5,0.25)};
 \addplot[only marks,mark=*,teal!80!black] coordinates {(1,0)};
 \draw[densely dashed,red] (axis cs:0.5,0) -- (axis cs:0.5,0.25);
 \node[font=\scriptsize,anchor=south west,align=left] at (axis cs:0.53,0.26) {$F'(\rho)=0$\\(fold)};
 \node[font=\scriptsize,anchor=north east,align=right] at (axis cs:1.20,-0.11) {$\kappa=0$\\(decoupling)};
\end{axis}
\begin{axis}[
 at={(source.east)},anchor=west,xshift=0.65cm,
 width=6.5cm,height=6.0cm,
 xmin=0,xmax=1.25,ymin=-0.3,ymax=1.2,
 axis lines=left,axis line style={-Latex},
 xlabel={$x=\lambda\rho/\kappa_0$},ylabel={$\kappa/\kappa_0$},
 xtick={0,0.5,1},xticklabels={$0$,$\frac12$,$1$},
 ytick={0,1},yticklabels={$0$,$1$},
 title={(b) Coupling $\kappa(\rho)$},title style={font=\small},
 tick label style={font=\scriptsize},clip=false]
 \addplot[blue,very thick,domain=0:1.22]{1-x};
 \draw[densely dashed,red] (axis cs:0.5,0) -- (axis cs:0.5,0.5);
 \addplot[only marks,mark=*,red] coordinates {(0.5,0.5)};
 \addplot[only marks,mark=*,teal!80!black] coordinates {(1,0)};
 \node[font=\scriptsize,anchor=south west] at (axis cs:0.51,0.52) {fold};
 \node[font=\scriptsize,anchor=south east] at (axis cs:0.99,0.10) {$\kappa=0$};
\end{axis}
\end{tikzpicture}
\caption{Linear dust model $\kappa=\kappa_0-\lambda\rho$, $\kappa_0,\lambda>0$. (a) Effective source $\lambda F/\kappa_0^2=x-x^2$: fold at $x=1/2$ (red), decoupling at $x=1$ (green). (b) Coupling $\kappa/\kappa_0=1-x$. The region $x>1$ ($F<0$) is the algebraic continuation of the map.}\label{fig:critical-surfaces}
\end{figure}
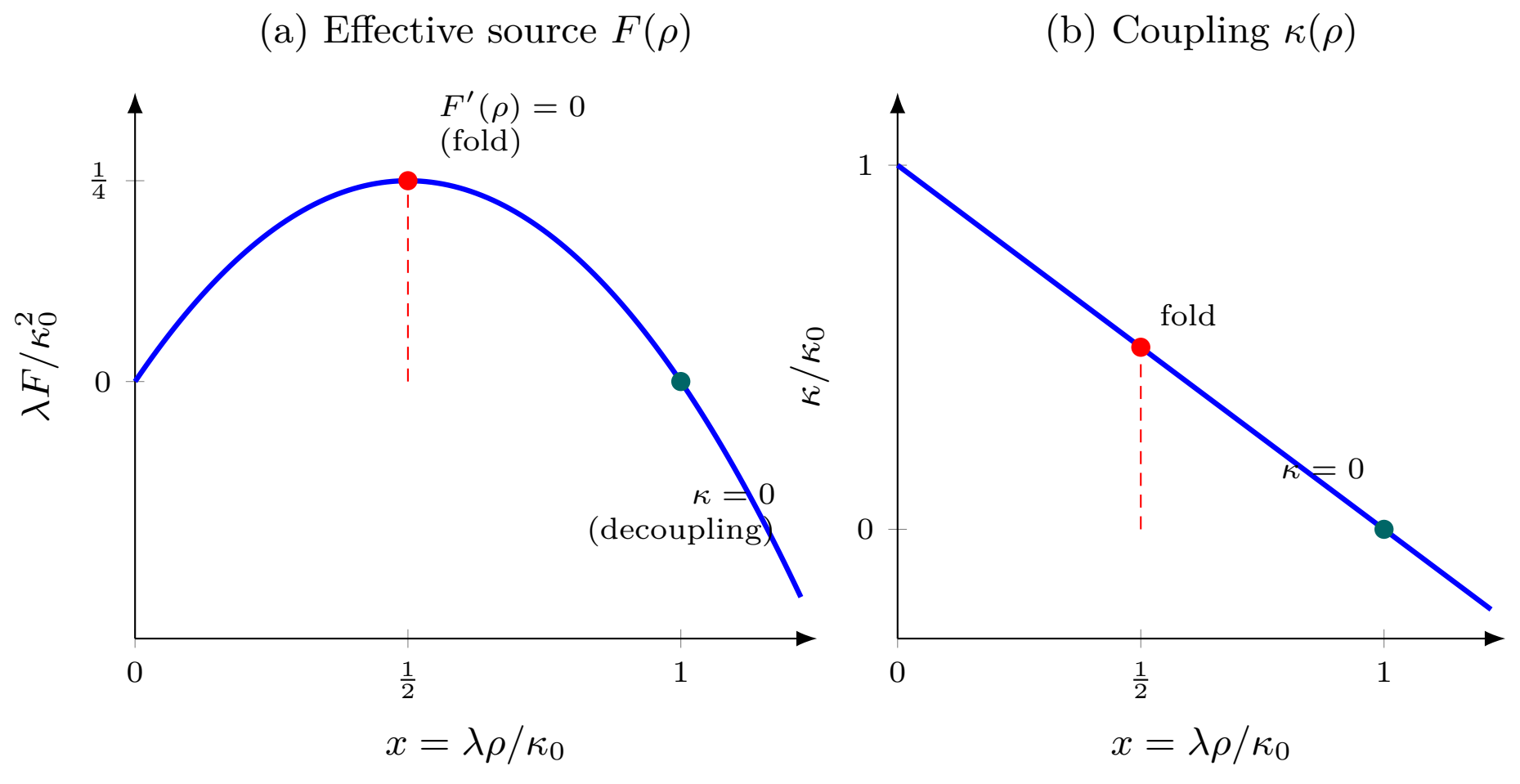

The action therefore locates the breakdown of the theory through its constraint structure: the loss of invertibility of the matter map coincides with an obstruction to consistent Hamiltonian evolution, not merely with a singular chart of the reduction. This is the canonical counterpart of item~3 of Proposition~\ref{prop:fold}. The geometric equations $E_{\mu\nu}=0$ do not involve $A$ or $C$, so the additional freedom of the adjoint sector at $F'=0$ cannot provide a continuation through the fold. The fold is also distinct from the decoupling surface $\kappa=0$ ($\rho=\rho_c$), where $F'=-\kappa_0\neq0$ and the reduction is regular (Fig.~\ref{fig:critical-surfaces}); the low-density branch reaches the fold first.

\subsection{Comparison with Loop Quantum Cosmology}\label{sec:lqccomp}
A Friedmann equation quadratic in the density, $H^2\propto\rho(1-\rho/\rho_c)$, is the hallmark of the effective dynamics of Loop Quantum Cosmology, where it replaces the big bang by a bounce at the critical density $\rho_c$ \cite{APS2006,Taveras2008,AshtekarSingh}. The same form arises in braneworlds with negative brane tension \cite{ShtanovSahni2003}, in mimetic gravity with limiting curvature \cite{ChamseddineMukhanov2017,Langlois2017}, and in energy--momentum-squared gravity \cite{EMSG,BoardBarrow2017}. In these frameworks it is usually taken as evidence of a bounce. The linear $\kappa(T)$ model shows that this conclusion requires more than the Friedmann equation: the same quadratic equation is compatible with a non-bouncing evolution once the conservation law is changed.

\begin{figure}[htbp]
\centering
\includegraphics[width=0.96\linewidth]{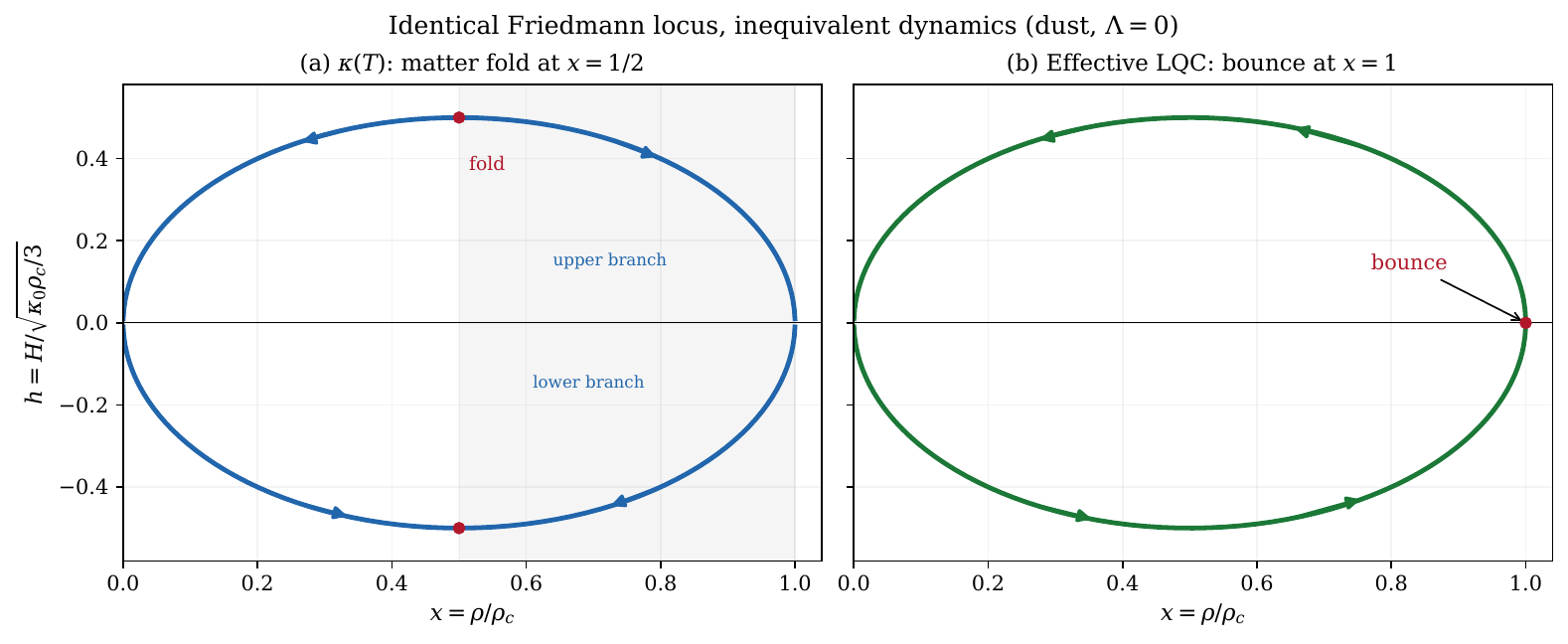}
\caption{Comparison of the linear $\kappa(T)$ dust cosmology and effective loop quantum cosmology. Despite their identical quadratic Friedmann equations, their conservation laws and time evolution differ: the former reaches a material fold, whereas the latter undergoes a bounce.}
\label{fig:kappa-lqc}
\end{figure}

In terms of $\rho_c:=\kappa_0/\lambda$, the Friedmann equation of the linear dust model reads
\begin{equation}\label{eq:lqc}
H^2=\frac{\kappa_0}{3}\rho\left(1-\frac{\rho}{\rho_c}\right)+\frac\Lambda3 ,
\end{equation}
This equation is already explicit in the original formulation of the theory, as $H^2=\frac{8\pi G}{3}\rho+\frac\Lambda3-\frac\lambda3\rho^2$ [Eq.~(19) of \cite{Perez}], where the bound $\rho\le8\pi G/\lambda$ was read off from $H^2\geq0$. Written in terms of $\rho_c$, it is for $\Lambda=0$ the effective Friedmann equation of Loop Quantum Cosmology; this identification and its consequences are new here. The critical density sits exactly on the decoupling surface $\kappa=0$. The two dynamics nevertheless differ:
\begin{center}\small
\begin{tabular}{@{}lll@{}}\toprule
 (dust, $\Lambda=0$) & linear $\kappa(T)$ & effective LQC\\\midrule
Friedmann & $H^2=F/3$ & $H^2=F/3$\\
conservation & $FV=$ const & $\rho V=$ const\\
Raychaudhuri & $\dot H=-\tfrac12F$ & $\dot H=-\tfrac12\rho F'(\rho)$\\
high density & fold at $\rho_c/2$ & bounce at $\rho_c$\\\bottomrule
\end{tabular}
\end{center}
The difference is not specific to the linear model. Figure~\ref{fig:kappa-lqc} illustrates the different evolution associated with the same quadratic Friedmann relation.

\begin{theorem}[No bounce with positive coupling]\label{prop:nobounce}
Consider $G_{\mu\nu}-\Lambda g_{\mu\nu}=\kappa(R,T)T_{\mu\nu}$ in flat FLRW with a barotropic fluid, $w>-1$, $\rho>0$. A bounce ($H=0$, $\dot H>0$) requires $\kappa<0$ and $\Lambda>0$ at the bounce.
\end{theorem}
\begin{proof}
Whatever the dependence of $\kappa$ on $R$ and $T$, the field equations are those of GR with the effective fluid $(\kappa\rho,\,w\kappa\rho)$. Hence $3H^2=\Lambda+\kappa\rho$ and $\dot H=-\frac12(1+w)\kappa\rho$. $H=0$ and $\dot H>0$ require $\kappa\rho=-\Lambda<0$.
\end{proof}
A bounce is therefore possible only in the repulsive phase $\kappa<0$ \cite{Teruel2026}. In the linear model this is the sector $M<0$ beyond the decoupling surface, where $3H^2=\Lambda-|M|/V$ gives, for $\Lambda>0$, the exact bouncing solution
\begin{equation}\label{eq:coshbounce}
V(t)=\frac{|M|}{\Lambda}\cosh^2\!\Bigl(\frac{\sqrt{3\Lambda}}{2}\,t\Bigr),
\end{equation}
with real matter on the branch $\rho>\rho_c$, where $F'<0$ and no fold occurs; at the bounce $F=-\Lambda$ and $\rho_b=\bigl[\kappa_0+\sqrt{\kappa_0^2+4\lambda\Lambda}\bigr]/(2\lambda)$. This bounce is driven by $\Lambda$ acting against a source of negative effective energy, with a scale set by $\Lambda$ rather than by a high density, and the sector has no solutions for $\Lambda\le0$. As shown in Section~\ref{sec:fold}, this sector is disconnected from the attractive low-density branch.

The ingredient that LQC adds can be identified. For conserved dust, define
\begin{equation}\label{eq:Sigma}
\Sigma_{\mu\nu}:=T_{\mu\nu}-\frac{2}{\rho_c}\,T\,T_{\mu\nu}+\frac{T^2}{\rho_c}\,g_{\mu\nu}.
\end{equation}

\begin{proposition}[Effective LQC as a pressure term]\label{prop:lqc}
For dust with $\nabla^\mu T_{\mu\nu}=0$, $\nabla^\mu\Sigma_{\mu\nu}=\frac{2\rho}{\rho_c}(\delta^\mu_\nu-u^\mu u_\nu)\nabla_\mu\rho$, which vanishes in FLRW. The equations $G_{\mu\nu}=\kappa_0\Sigma_{\mu\nu}$ then reproduce the effective LQC Friedmann and Raychaudhuri equations exactly, with effective density $\rho(1-\rho/\rho_c)$ and pressure $-\rho^2/\rho_c$.
\end{proposition}
\begin{proof}
Using $\nabla^\mu T_{\mu\nu}=0$, $\nabla^\mu\Sigma_{\mu\nu}=\frac{2}{\rho_c}\bigl(T\nabla_\nu T-T_{\mu\nu}\nabla^\mu T\bigr)$, which for $T_{\mu\nu}=\rho u_\mu u_\nu$ gives the stated projected gradient. Writing $\Sigma_{\mu\nu}=(\rho_{\rm e}+p_{\rm e})u_\mu u_\nu-p_{\rm e}g_{\mu\nu}$ gives $\rho_{\rm e}=\rho(1-\rho/\rho_c)$ and $p_{\rm e}=-\rho^2/\rho_c$. Then $3H^2=\kappa_0\rho_{\rm e}$ and $\dot H=-\frac{\kappa_0}2(\rho_{\rm e}+p_{\rm e})=-\frac{\kappa_0}2\rho\bigl(1-2\rho/\rho_c\bigr)$.
\end{proof}

Since $\kappa(T)T_{\mu\nu}=\kappa_0(T_{\mu\nu}-T\,T_{\mu\nu}/\rho_c)$ for dust, effective LQC equals the linear $\kappa(T)$ model plus the isotropic pressure $-\rho^2/\rho_c$. That pressure is what allows ordinary conservation, violates the null energy condition and produces the bounce. Terms of the form $T\,T_{\mu\nu}$ and $T^2g_{\mu\nu}$ appear in energy--momentum-squared gravity, where bounces with a maximal density have been reported \cite{EMSG}.

\paragraph{The missing pressure is the work of the medium.} The pressure has a precise origin. The fluid with the same energy function of the \emph{physical} number, $\mathcal L_c=-\sqrt{|g|}\,(\kappa/\kappa_0)\rho(n)$, is an ordinary Lagrangian fluid that conserves $n$. Its pressure is
\begin{equation}\label{eq:pcons}
p_c=n\frac{d\varepsilon}{dn}-\varepsilon=\frac{\kappa}{\kappa_0}\,p+\rho\,\frac{d(\kappa/\kappa_0)}{d\ln n}.
\end{equation}
The second term is the compression work of the induced sector of Section~\ref{sec:medium}: with energy $\chi_g\rho$, its thermodynamic pressure is $n^2\,d(\chi_g\rho/n)/dn=\chi_g p+\rho\,d\chi_g/d\ln n$. The field equations of $\kappa(R,T)$ keep $\chi_g p$ and drop the work term, which is why they conserve $N$ rather than $n$. For linear dust the work term is $-\rho^2/\rho_c$, and $\mathcal L_c$ has stress tensor $\Sigma_{\mu\nu}$:
\begin{itemize}
\item \emph{$\kappa(R,T)$ and effective LQC are the same energy function with different conserved numbers}: the dressed number gives the fold, the physical number gives the bounce.
\item Since $\Sigma_{\mu\nu}$ is the stress tensor of a Lagrangian fluid, it is conserved beyond FLRW whenever that fluid obeys its own Euler equations, which include the pressure gradient. It is not geodesic dust. Its sound speed is
\[
c_s^2=\frac{dp_c}{d\varepsilon}=-\frac{2\rho/\rho_c}{1-2\rho/\rho_c},
\]
negative below $\rho_c/2$, divergent at $\rho_c/2$ and superluminal for all $\rho>\rho_c/2$. The bounce of the conservative completion is therefore a homogeneous statement: its perturbations meet the fold density again, as a divergence of the sound speed. This sound speed belongs to the conservative fluid realization
constructed here. The homogeneous correspondence does not
identify its perturbations with those of LQC.
\item A heavy auxiliary scalar coupled to the trace, $\mathcal L=-\sqrt{|g|}\,[\rho-\alpha\varphi T+\frac12M^2\varphi^2]$, gives $\varepsilon=\rho-\beta T^2$ with $\beta=\alpha^2/2M^2$ once $\varphi$ is eliminated. For dust this is exactly $\mathcal L_c$ with $\rho_c=1/\beta$: the medium that returns its work can be realized explicitly, and it yields effective LQC rather than $\kappa(R,T)$.
\end{itemize}

\paragraph{The fold and the LQC signature change.} The density $\rho_c/2$ has a known meaning in LQC. In the anomaly-free treatment of holonomy-corrected perturbations, the hypersurface-deformation algebra is deformed by
\begin{equation}\label{eq:Omega}
\Omega=1-\frac{2\rho}{\rho_c},
\end{equation}
which multiplies the spatial Laplacian. Its vanishing at $\rho_c/2$ marks a change from Lorentzian to Euclidean effective signature, which persists up to the bounce \cite{Cailleteau2012,Mielczarek2012}. Since $\Omega=F'(\rho)/\kappa_0$ for the shared energy function, the material fold of $\kappa(R,T)$ is located exactly where LQC changes signature. The coincidence has a structural origin: both quantities are the derivative of $\rho(1-\rho/\rho_c)$. The two theories differ in what they do there. LQC continues through a Euclidean phase to a bounce at $\rho_c$. In $\kappa(R,T)$ the same point is a degeneracy of the constitutive map and of the reduced constraint matrix, where the evolution ends. The correspondence is specific to dust: the LQC correction is universal in $\rho$, whereas $\kappa(T)$ responds to the trace and does not act on radiation.

\section{Quantization and the adjoint field}\label{sec:quantum}
An action is also the usual starting point for quantization, and this was one of the motivations for seeking one for $\kappa(R,T)$. Here the structure of Chen's action, linear in the multiplier, determines the outcome. In the reduced dust model the constraint is linear in the momenta and generates the classical flow, so its quantization yields a transport equation that can be solved exactly. In the covariant vacuum theory the quadratic action pairs the metric with the multiplier, so the free propagator is purely mixed. Both results are features of the response-field family to which the action belongs. We present them as such. They are established for the reduced minisuperspace with a chosen ordering, and for the bare vacuum quadratic theory, respectively; they do not show that every covariant quantization of Chen's action has the same structure.

\subsection{Dust: transport quantization}
For $w=0$ set $p_V=p_q/(3V)$ and $P:=p_C$. Then \eqref{eq:Hred} becomes
\begin{equation}\label{eq:Hdust}
\cH_{\rm dust}=-\frac32Pp_V+C\left(\frac{3P^2}{4V}+\Lambda V\right).
\end{equation}
Taking $P$ as configuration variable, $p_P=-C$, and
\begin{equation}\label{eq:transport}
\cH_{\rm dust}=X^Vp_V+X^Pp_P,\qquad X^V=-\frac32P,\qquad X^P=-\left(\frac{3P^2}{4V}+\Lambda V\right),
\end{equation}
which is the cotangent lift of the vector field $X$ generating the Friedmann and acceleration equations ($P=-2VH$). The function
\begin{equation}
M(V,P)=\frac{3P^2}{4V}-\Lambda V=VF
\end{equation}
satisfies $XM=0$; it is the Bianchi invariant \eqref{eq:bianchi}. With $p_V=-i\hbar\partial_V$, $p_P=-i\hbar\partial_P$ and the symmetric ordering $\widehat\cH=-i\hbar(X+\frac12\Div X)$, $\Div X=-3P/(2V)$, the constraint is
\begin{equation}\label{eq:PDE}
\left[-\frac{3P}{2}\partial_V-\left(\frac{3P^2}{4V}+\Lambda V\right)\partial_P-\frac{3P}{4V}\right]\Psi=0,
\end{equation}
with general local solution
\begin{equation}\label{eq:sol}
\Psi(V,P)=V^{-1/2}f\!\left(\frac{3P^2}{4V}-\Lambda V\right).
\end{equation}
The current $J^i=X^i|\Psi|^2$ is conserved. On a section $V=V_0$ of the expanding branch $P<0$,
\begin{equation}\label{eq:norm}
\lVert\Psi\rVert^2_{V_0}=\int_{P<0}\dd P\,X^V|\Psi(V_0,P)|^2=\int_{\mathcal I_{V_0}}\dd M\,|f(M)|^2,
\end{equation}
which is positive and independent of the section as long as no characteristics leave through a boundary. In the linear model the fold $V=4\lambda M/\kappa_0^2$ is such a boundary: $X(V-4\lambda M/\kappa_0^2)=XV\neq0$, so it is transverse to the flow.

Equation \eqref{eq:PDE} is the Liouville equation for half-densities, and $\hbar$ drops out of it entirely. In this reduced model the quantization is therefore of Koopman--von Neumann type \cite{KvN}: it propagates distributions over classical trajectories, labelled by $M$. Since the equation is first order, $|\Psi|^2$ is itself transported along the classical flow, and the dynamics generates no interference; a classical statistical reading requires restricting the observables to functions on the classical phase space. In this sense the reduced quantization gives a consistent statistical description of the classical cosmological dynamics, not a quantum theory of gravity.

\subsection{Vacuum quadratic theory}
Around $\bar g_{\mu\nu}=\eta_{\mu\nu}$, $\bar a^{\mu\nu}=0$, $T_{\mu\nu}=0$, $\Lambda=0$, the quadratic action is exactly
\begin{equation}\label{eq:S2}
S^{(2)}=\int\dd^4x\,b^{\mu\nu}G^{(1)}_{\mu\nu}[h]=:\langle b,\cE h\rangle ,
\end{equation}
with $\cE$ formally self-adjoint. In the field vector $(h,b)$ the Hessian and its gauge-fixed inverse are
\begin{equation}\label{eq:K}
\mathbb K=\begin{pmatrix}0&\cE^\dagger\\\cE&0\end{pmatrix},\qquad
\mathbb K^{-1}=\begin{pmatrix}0&\cE^{-1}\\(\cE^\dagger)^{-1}&0\end{pmatrix},
\end{equation}
so $\langle hh\rangle_0=\langle bb\rangle_0=0$ and $\langle hb\rangle_0=i\hbar\,\cE^{-1}$. The inverse is understood after gauge fixing and with a specified
Green-function prescription. It determines the bare quadratic
kernel; state-dependent correlations require additional input. Both $h$ and $b$ have linearized diffeomorphism redundancies; $\delta b_{\mu\nu}=\partial_\mu\zeta_\nu+\partial_\nu\zeta_\mu$ is a symmetry by the linearized Bianchi identity. On the transverse-traceless sector, $\cE=-\Box/2$, and with $u=(h+b)/\sqrt2$, $v=(h-b)/\sqrt2$,
\begin{equation}\label{eq:diag}
S^{(2)}_{\rm TT}=\tfrac12\langle u,\cE u\rangle-\tfrac12\langle v,\cE v\rangle .
\end{equation}
The adjoint sector doubles the tensor modes with an indefinite kinetic form. The zero-adjoint sector is a consistent classical truncation,
but this does not establish stability of the full extended
system. If the diagonalized sectors are interpreted as
independent ordinary quantum fields, their opposite kinetic
signs obstruct a simultaneous positive-energy and positive-norm
interpretation. A response-field formulation requires a separate
specification of contour, state, measure and physical observables.

This is the standard structure of response-field actions. Integrating the multiplier gives $\int\mathcal Db\,e^{\frac i\hbar\langle b,E\rangle}\propto\delta[E]$, which localizes the path integral on classical solutions, as in the classical path integral \cite{Gozzi}. In a background with matter, \eqref{eq:S2} acquires a $b$--$\chi$ block through $\kappa\,\delta T_{\mu\nu}$, which we have not included.

\subsection{Physical meaning of the adjoint field}\label{sec:adjoint}
The adjoint field never acts on the geometry: $E_{\mu\nu}=0$ does not contain $a^{\mu\nu}$, and $a=0$ is the physical limit. Its content is in the response of the theory, which can be made precise in three ways.

\paragraph{\textbf{Response identity.}} Add an external source to the equations, $S_J=\int\sqrt{|g|}\,a^{\mu\nu}(E_{\mu\nu}-J_{\mu\nu})$, and consider the generating functional $Z[J]=\int\mathcal Dg\,\mathcal Da\,e^{\frac i\hbar S_J}$. Integration over $a$ formally imposes $E=J$, with the corresponding Jacobian and boundary prescription. For normalized expectation values and a
source-independent measure and domain,
differentiating
$\langle g_{\alpha\beta}(x)\rangle_J$
with respect to $J_{\mu\nu}(y)$ gives
\begin{equation}\label{eq:MSR}
\bigl\langle g_{\alpha\beta}(x)\,\sqrt{|g(y)|}\,a^{\mu\nu}(y)\bigr\rangle_{\rm c}=i\hbar\,\frac{\delta\langle g_{\alpha\beta}(x)\rangle_J}{\delta J_{\mu\nu}(y)} ,
\end{equation}
where the subscript denotes the connected correlator. The mixed correlator is the linear response of the metric to a perturbation of the field equations. In the vacuum quadratic theory, $\cE h=J$ gives $\delta h/\delta J=\cE^{-1}$, which is the mixed propagator \eqref{eq:K}. The adjoint field is thus the response field of the Martin--Siggia--Rose construction \cite{MSR}: it is not measured, but it carries the Green function.

\paragraph{\textbf{Adjoint variables as conserved sensitivities.}} In the reduced dust model the constraint \eqref{eq:transport} has the form $\cH=X^i(x)p_i$, with configuration $x=(V,P)$ and adjoint momenta $p=(p_V,p_P)$, $p_P=-C$. In proper time ($N=1$) Hamilton's equations read
\begin{equation}
\dot x^i=X^i,\qquad \dot p_i=-\partial_iX^j\,p_j ,
\end{equation}
so the adjoint variables obey the transposed linearized flow. For any perturbation $\delta x$ of a classical trajectory, $\delta\dot x^i=\partial_jX^i\delta x^j$, and
\begin{equation}\label{eq:pairing}
\frac{\dd}{\dd t}\bigl(p_i\,\delta x^i\bigr)=0 .
\end{equation}
The adjoint field is therefore the conserved covector dual to perturbations. If $p_i(t_f)=\partial\mathcal O/\partial x^i$ for an observable $\mathcal O$ evaluated at $t_f$, then $p_i(t)=\partial\mathcal O(x(t_f))/\partial x^i(t)$ is the sensitivity of that observable to a perturbation at time $t$. Every conserved quantity $Q$ gives an exact adjoint solution $p=\nabla Q$, because differentiating $X^j\partial_jQ=0$ yields $\frac{\dd}{\dd t}\partial_iQ=-\partial_iX^j\partial_jQ$; it satisfies the constraint $X\cdot\nabla Q=0$, and $p_i\delta x^i=\delta Q$. For the Bianchi invariant $M$ of \eqref{eq:bianchi}, using $P=-2VH$,
\begin{equation}\label{eq:adjointM}
C=-\partial_PM=3H,\qquad p_V=\partial_VM=-(3H^2+\Lambda) .
\end{equation}
The adjoint configuration $C=3H$ measures how a perturbation of the geometry changes the effective mass of the universe. The matter density depends on that label through $\rho=F^{-1}(M/V)$, with sensitivity
\begin{equation}\label{eq:rhosens}
\frac{\partial\rho}{\partial M}=\frac{1}{V F'(\rho)} ,
\end{equation}
which diverges at the fold. The same factor $F'$ multiplies $A-wC$ in the constraint $Y$, so the determination of the adjoint variable $A$ is lost exactly where the response of the matter to the geometry diverges.

\paragraph{\textbf{Doubled actions and the Helmholtz condition.}} For a specified system of local equation densities
$\mathcal E_i[\phi]$, formal self-adjointness of
its linearization is the Helmholtz condition:
under the standard local assumptions, it is
necessary and locally sufficient for
$\mathcal E_i=\delta S/\delta\phi^i$
\cite{Olver}. The criterion applies in the chosen
field variables and includes the appropriate
density factors and index conventions. It does
not exclude equivalent variational formulations
obtained with multipliers or additional fields.
When it holds, a closed-time-path action $S[\phi_+]-S[\phi_-]$ with $\phi_\pm=\phi\pm\frac12 a$ expands as
\begin{equation}\label{eq:CTPexpansion}
S[\phi+\tfrac12a]-S[\phi-\tfrac12a]=\int a\,\frac{\delta S}{\delta\phi}+O(a^3),
\end{equation}
so the leading term is $\int a^i\mathcal E_i$.
Chen's action shares this multiplier-times-equations
structure through the pairing
$\int\sqrt{|g|}\,a^{\mu\nu}E_{\mu\nu}$, and remains
available without requiring the prescribed equations
to satisfy the Helmholtz condition. This structural
correspondence does not by itself identify the
multiplier with a metric difference field. The non-conservation of $T_{\mu\nu}$ does not by itself make the system non-variational. For $\kappa=\kappa(T)$ and an isentropic barotropic
fluid, wherever the dressed density gives a regular
local constitutive parametrization, the energy
function $\varepsilon(N)=\kappa\rho/\kappa_0$
satisfies
$N\,d\varepsilon/dN-\varepsilon=\kappa p/\kappa_0$,
by \eqref{eq:dressed}. It therefore admits an
ordinary effective perfect-fluid realization,
with the physical density reconstructed on a
chosen inverse branch. This does not establish
equivalence of the full covariant theories.
Theorem~\ref{thm:universal} establishes the
corresponding canonical universality in the
constant-$w$ homogeneous reduction.
This effective-fluid description requires a local
constitutive relation and uses dressed rather than
physical matter variables. At a material fold,
the map $n\mapsto N$ loses local invertibility,
so recovering the physical density requires a
choice of branch; the effective-fluid action itself
need not become singular. At critical coupling,
$N=0$ can coexist with nonzero physical matter,
whereas in the repulsive dust phase $N<0$ is a
signed density rather than a positive particle count.
These features limit the reconstruction and its
physical interpretation, but do not establish the
nonexistence of ordinary variational formulations.
Chen's adjoint action enforces the original field
equations wherever they are defined, without
dividing by $\kappa$ or inverting the matter map.
Its formulation therefore remains available at
folds and critical surfaces, although this does
not guarantee regular evolution through them or
exclude alternative variational formulations with
additional fields or constraints.

\paragraph{\textbf{Open-system interpretation.}} These properties suggest a more physical interpretation. In the Keldysh basis, the term linear in the difference field generates the mean equations
of motion \cite{Schwinger,Keldysh,Galley}.
Retaining only this term defines the deterministic
response truncation considered here. A classical
statistical limit can also retain quadratic
difference-field terms associated with noise;
classicality alone does not justify discarding them.
In this interpretation $a^{\mu\nu}$ is the difference component of the metric, and $\kappa(R,T)$ is the classical effective dynamics of gravity coupled to an environment. Section~\ref{sec:medium} suggests a candidate: an environment whose induced stress \eqref{eq:Tenv} is a local function of the matter. This is consistent with the exchange of energy
encoded in $\Gamma=-\dot\kappa/\kappa$. The structural analogy does not by itself show that Chen's action is the effective action of an open system; that would require identifying the environment, an influence functional and the corresponding positivity conditions. An open-system completion could add terms quadratic in the difference field, $\int a^{\mu\nu}\mathcal N_{\mu\nu\rho\sigma}a^{\rho\sigma}$, describing stochastic sources as in stochastic semiclassical gravity \cite{HuVerdaguer}. The relation of such a noise kernel to the exchange rate $\Gamma$ would depend on the microscopic environment and its state, since different environments can produce the same mean equation with different noise.
A subsequent construction provides an explicit local covariant
parent action with dust variables, an algebraic density field
and an internal oscillator \cite{ChenMedium2026}.
With a specified zero-oscillator preparation, its smooth
flat-FLRW stationary histories reproduce the target dust
equations. Its selected classical response reduction identifies
the adjoint multiplier with a rescaled inverse-metric difference,
$a^{\mu\nu}=\eta_g^{\mu\nu}/(2\kappa_0)$.
In that construction the medium field is algebraic and its elimination reparametrizes the energy function of the dust, so the physical density is not yet identified operationally independently of the total source; a medium with dynamics of its own remains to be constructed.
This provides a formal realization of the proposed interpretation in that sector;
inhomogeneous recovery, propagating-mode stability and a complete
finite-width quantum influence functional remain open.

\section{Conclusions}\label{sec:conclusions}
Chen's extended action answers the objection that
$\kappa(R,T)$ gravity has no variational formulation.
It does so through a local adjoint action with exact
projected recovery of the prescribed field equations.
Its adjoint structure is generic, but its application
here yields concrete canonical and response results.
An ordinary effective-fluid description is available
where the constitutive reconstruction is regular;
material folds obstruct smooth reconstruction of
the original matter variables without necessarily
making the effective action singular.
The induced-source decomposition admits a concrete
medium interpretation: Chen's subsequent construction
provides an explicit local covariant parent model
whose prepared homogeneous dust sector reproduces
the target equations and whose selected response
reduction recovers the adjoint structure
\cite{ChenMedium2026}. This provides a formal realization
in that sector, while its extension to general
inhomogeneous matter remains open.
In flat FLRW with a prescribed constant-$w$
fluid, the results are:
\begin{itemize}
\item the reduced Hamiltonian is universal in $\kappa$, and the coupling survives only in a second-class constraint;
\item matter is generically non-conservative, with $\Gamma=-\dot\kappa/\kappa$; for a homogeneous constant-$w$ fluid with $\rho>0$ and $w\neq-1,1/3$, ordinary conservation is incompatible with the modified balance law on a non-static FLRW interval where $\kappa_T\neq0$;
\item for a linear coupling, the degeneration of that constraint coincides with a material fold at $\rho_c/2$, which obstructs consistent Hamiltonian evolution: analytic geometry, square-root singularity of the matter, no continuation of an expanding solution to earlier times for $\Lambda\geq0$, and a two-valued reconstruction of the matter from the fold;
\item the balance law is time-symmetric, and the theory is exactly general relativity with matter plus an induced source of susceptibility $\chi_g=\kappa/\kappa_0-1$; critical surfaces are total screening, the repulsive phase is over-screening, and the fold is a saturation-like degeneracy of the constitutive map, $\det\mathcal K=\kappa^9F'$ for dust;
\item the theory reproduces the effective LQC Friedmann equation but cannot bounce with $\kappa>0$. For dust, effective LQC is the same energy function with the physical rather than the dressed number conserved; the missing pressure $-\rho^2/\rho_c$ is the compression work of the induced medium, and the fold sits exactly at the density $\rho_c/2$ where LQC changes signature in the deformed-algebra approach; bounces exist only in the repulsive phase $\kappa<0$, disconnected from the attractive branch in homogeneous cosmology, and the ingredient missing for an LQC-type bounce is the pressure $-\rho^2/\rho_c$;
\item in the reduced dust model the quantized constraint is a transport equation of Koopman--von Neumann type, solved exactly, and the bare vacuum propagator is purely mixed;
\item the adjoint field is the response variable of the theory and suggests an interpretation as the difference field of an open-system (Schwinger--Keldysh) action, consistent with the energy exchange that characterizes $\kappa(R,T)$; an open-system completion could add stochastic geometric fluctuations, whose relation to $\Gamma$ would depend on the environment.
\end{itemize}
The results rely on a prescribed matter source, and the fold marks the point where that description fails. The natural next steps are a matter sector with its own dynamics, to determine what such matter actually does as it approaches the fold, and the behaviour of inhomogeneous perturbations there, which could make the fold dynamically unreachable.

\paragraph{\textbf{Outlook: gravity in a polarizable medium.}} The medium form \eqref{eq:medium} suggests a programme for which a first formal realization is now available. Chen's subsequent construction supplies a local covariant parent action whose prepared homogeneous dust sector recovers the target equations, together with a selected classical response reduction yielding the adjoint structure
\cite{ChenMedium2026}.
The next step is to extend this recovery beyond
homogeneous dust and to establish a controlled
influence-functional description. In particular,
one should determine whether, for suitable
environmental fields $\phi$ and a specified
initial state, the closed-time-path integral
\begin{equation}
e^{iS_{\rm eff}[g_+,g_-]}=\int\mathcal D\phi_+\mathcal D\phi_-\,e^{iS[g_+,\phi_+]-iS[g_-,\phi_-]}
\end{equation}
yields the mean equation $G_{\mu\nu}-\Lambda g_{\mu\nu}=\kappa(R,T)T_{\mu\nu}$, and then to check whether, in the Keldysh basis, the term linear in the difference field coincides with Chen's action. This is more demanding than reproducing $T^{\rm env}=\chi T$ algebraically. Not every local constitutive relation follows from integrating out auxiliary fields with a local, covariant microscopic action, and the diffeomorphism invariance of the complete system imposes conservation identities that the medium must respect. Reproducing exactly $\kappa(T)T_{\mu\nu}$, without additional tensor structures, may therefore be difficult; this is precisely what makes the question informative. Such an effective action would generally have the form
\begin{equation}
S_{\rm eff}=\int a^{\mu\nu}E_{\mu\nu}+\frac i2\int a^{\mu\nu}\mathcal N_{\mu\nu\rho\sigma}a^{\rho\sigma}+\cdots,
\end{equation}
and the locality of the mean equation does not force $\mathcal N=0$. Whether Chen's action is complete at this order, and which geometric fluctuations accompany the exchange of energy with the medium, depend on the microscopic model and on the state of the medium, and are left for future work.

\section*{Acknowledgments}
I thank J.~Chen for correspondence and for checking the description of his construction. AI-based assistants (Claude, Anthropic; ChatGPT, OpenAI) were used for symbolic and numerical checks, literature searches and language editing. The author designed the study, reviewed all results and takes full responsibility for the content.


\begin{thebibliography}{99}
\bibitem{Perez} G.~R.~P\'erez Teruel, \emph{$\kappa(R,T)$ gravity}, Eur. Phys. J. C \textbf{78}, 660 (2018).
\bibitem{AhmedPradhan2020} N.~Ahmed and A.~Pradhan, \emph{Probing $\kappa(R,T)$ cosmology via empirical approach}, arXiv:2002.03798 [gr-qc].
\bibitem{AhmedPradhan2022} N.~Ahmed and A.~Pradhan, \emph{Probing cosmic acceleration in $\kappa(R,T)$ gravity}, Indian J. Phys. \textbf{96}, 301 (2022).
\bibitem{Dixit2022} A.~Dixit, A.~Pradhan and R.~Chaubey, \emph{Cosmological scenario in $\kappa(R,T)$ gravity}, Int. J. Geom. Methods Mod. Phys. \textbf{19}, 2250013 (2022).
\bibitem{Dixit2023} A.~Dixit, S.~Gupta, A.~Pradhan and A.~Beesham, \emph{Thermodynamics of the acceleration of the universe in the $\kappa(R,T)$ gravity model}, Symmetry \textbf{15}, 549 (2023).
\bibitem{Teruel2022} G.~R.~P\'erez Teruel, K.~Newton Singh, F.~Rahaman and T.~Chowdhury, \emph{Possible existence of stable compact stars in $\kappa(R,T)$ gravity}, Int. J. Mod. Phys. A \textbf{37}, 2250194 (2022).
\bibitem{TaserDogru2023} D.~Ta\c{s}er and S.~S.~Do\u{g}ru, \emph{Compact stars in $\kappa(R,T)$ gravity}, Astrophys. Space Sci. \textbf{368}, 49 (2023).
\bibitem{Sarkar2022} S.~Sarkar, N.~Sarkar, F.~Rahaman and Y.~Aditya, \emph{Wormholes in $\kappa(R,T)$ gravity}, arXiv:2207.12403 [gr-qc].
\bibitem{Singh2024} K.~N.~Singh, G.~R.~P\'erez Teruel, S.~K.~Maurya, T.~Chowdhury and F.~Rahaman, \emph{Conservative wormholes in generalized $\kappa(R,T)$ function}, J. High Energy Astrophys. \textbf{44}, 132 (2024), arXiv:2403.19733 [gr-qc].
\bibitem{Teruel2024} G.~R.~P\'erez Teruel, K.~Newton Singh, F.~Rahaman, T.~Chowdhury and M.~Mondal, \emph{Testing of $\kappa(R,T)$-gravity through gravastar configurations}, Phys. Dark Universe \textbf{43}, 101404 (2024).
\bibitem{Teruel2026} G.~R.~P\'erez Teruel, \emph{Critical coupling surfaces in $\kappa(R,T)$ gravity: regularity, gravitational screening, and phase transitions}, arXiv:2606.11942 [gr-qc].
\bibitem{Chen} J.~Chen, \emph{A Local Extended Action for $\kappa(R,T)$ Gravity with Exact Projected Recovery}, Zenodo, version 1.0.0 (28 October 2026), doi:\href{https://doi.org/10.5281/zenodo.23722034}{10.5281/zenodo.23722034}.
\bibitem{AthertonHomsy} R.~W.~Atherton and G.~M.~Homsy, \emph{On the existence and formulation of variational principles for nonlinear differential equations}, Stud. Appl. Math. \textbf{54}, 31 (1975), doi:10.1002/sapm197554131.
\bibitem{Ibragimov} N.~H.~Ibragimov, \emph{Integrating factors, adjoint equations and Lagrangians}, J. Math. Anal. Appl. \textbf{318}, 742 (2006), doi:10.1016/j.jmaa.2005.11.012.
\bibitem{Peng} L.~Peng, \emph{A modified formal Lagrangian formulation for general differential equations}, Japan J. Indust. Appl. Math. \textbf{39}, 573 (2022), arXiv:2009.04102.
\bibitem{ChenTutorial} J.~Chen, \emph{Reading Action and Path Integrals from Local Evolution}, Zenodo, version 3 (2026), doi:10.5281/zenodo.23681626.
\bibitem{ChenRecon} J.~Chen, \emph{Reconstructing Actions by Solving Compatibility Conditions for Unknown Multipliers}, Zenodo (2026), doi:10.5281/zenodo.23683535.
\bibitem{PerezPena} G.~R.~P\'erez Teruel and A.~Pe\~na Pe\~na, \emph{A modified gravitational theory of the matter sector of the type $\phi(R,T)\mathcal L_m$}, arXiv:2511.06811v2 [gr-qc] (2026).
\bibitem{Bateman} H.~Bateman, \emph{On dissipative systems and related variational principles}, Phys. Rev. \textbf{38}, 815 (1931).
\bibitem{MSR} P.~C.~Martin, E.~D.~Siggia and H.~A.~Rose, \emph{Statistical dynamics of classical systems}, Phys. Rev. A \textbf{8}, 423 (1973).
\bibitem{Gozzi} E.~Gozzi, M.~Reuter and W.~D.~Thacker, \emph{Hidden BRS invariance in classical mechanics. II}, Phys. Rev. D \textbf{40}, 3363 (1989).
\bibitem{Galley} C.~R.~Galley, \emph{Classical mechanics of nonconservative systems}, Phys. Rev. Lett. \textbf{110}, 174301 (2013).
\bibitem{Harko} T.~Harko, F.~S.~N.~Lobo, S.~Nojiri and S.~D.~Odintsov, \emph{$f(R,T)$ gravity}, Phys. Rev. D \textbf{84}, 024020 (2011).
\bibitem{Prigogine} I.~Prigogine, J.~Geheniau, E.~Gunzig and P.~Nardone, \emph{Thermodynamics and cosmology}, Gen. Relativ. Gravit. \textbf{21}, 767 (1989).
\bibitem{APS2006} A.~Ashtekar, T.~Pawlowski and P.~Singh, \emph{Quantum nature of the big bang: improved dynamics}, Phys. Rev. D \textbf{74}, 084003 (2006).
\bibitem{Taveras2008} V.~Taveras, \emph{Corrections to the Friedmann equations from loop quantum gravity for a universe with a free scalar field}, Phys. Rev. D \textbf{78}, 064072 (2008), arXiv:0807.3325.
\bibitem{ShtanovSahni2003} Y.~Shtanov and V.~Sahni, \emph{Bouncing braneworlds}, Phys. Lett. B \textbf{557}, 1 (2003), arXiv:gr-qc/0208047.
\bibitem{ChamseddineMukhanov2017} A.~H.~Chamseddine and V.~Mukhanov, \emph{Resolving cosmological singularities}, JCAP \textbf{03}, 009 (2017), arXiv:1612.05860.
\bibitem{Langlois2017} D.~Langlois, H.~Liu, K.~Noui and E.~Wilson-Ewing, \emph{Effective Loop Quantum Cosmology as a higher-derivative scalar-tensor theory}, Class. Quantum Grav. \textbf{34}, 225004 (2017), arXiv:1703.10812.
\bibitem{BoardBarrow2017} C.~V.~R.~Board and J.~D.~Barrow, \emph{Cosmological models in energy-momentum-squared gravity}, Phys. Rev. D \textbf{96}, 123517 (2017), arXiv:1709.09501.
\bibitem{AshtekarSingh} A.~Ashtekar and P.~Singh, \emph{Loop quantum cosmology: a status report}, Class. Quantum Grav. \textbf{28}, 213001 (2011).
\bibitem{EMSG} M.~Roshan and F.~Shojai, \emph{Energy-momentum squared gravity}, Phys. Rev. D \textbf{94}, 044002 (2016).
\bibitem{Schwinger} J.~Schwinger, \emph{Brownian motion of a quantum oscillator}, J. Math. Phys. \textbf{2}, 407 (1961).
\bibitem{Keldysh} L.~V.~Keldysh, \emph{Diagram technique for nonequilibrium processes}, Zh. Eksp. Teor. Fiz. \textbf{47}, 1515 (1964) [Sov. Phys. JETP \textbf{20}, 1018 (1965)].
\bibitem{HuVerdaguer} B.~L.~Hu and E.~Verdaguer, \emph{Stochastic gravity: theory and applications}, Living Rev. Relativ. \textbf{11}, 3 (2008).
\bibitem{Olver} P.~J.~Olver, \emph{Applications of Lie Groups to Differential Equations}, 2nd ed., Graduate Texts in Mathematics \textbf{107} (Springer, New York, 1993).
\bibitem{KvN} B.~O.~Koopman, Proc. Natl. Acad. Sci. USA \textbf{17}, 315 (1931); J.~von Neumann, Ann. Math. \textbf{33}, 587 (1932).
\bibitem{Cailleteau2012} T.~Cailleteau, J.~Mielczarek, A.~Barrau and J.~Grain, \emph{Anomaly-free scalar perturbations with holonomy corrections in loop quantum cosmology}, Class. Quantum Grav. \textbf{29}, 095010 (2012), arXiv:1111.3535.
\bibitem{Mielczarek2012} J.~Mielczarek, \emph{Signature change in loop quantum cosmology}, Springer Proc. Phys. \textbf{157}, 555 (2014), doi:10.1007/978-3-319-06761-2\_77, arXiv:1207.4657.
\bibitem{Rastall} P.~Rastall, \emph{Generalization of the Einstein theory}, Phys. Rev. D \textbf{6}, 3357 (1972).
\bibitem{Visser2018} M.~Visser, \emph{Rastall gravity is equivalent to Einstein gravity}, Phys. Lett. B \textbf{782}, 83 (2018), arXiv:1711.11500.
\bibitem{Darabi2018} F.~Darabi, H.~Moradpour, I.~Licata, Y.~Heydarzade and C.~Corda, \emph{Einstein and Rastall theories of gravitation in comparison}, Eur. Phys. J. C \textbf{78}, 25 (2018), arXiv:1712.09307.
\bibitem{ChenMedium2026}
J.~Chen,
\emph{A Local Medium Action for Density-Dependent Dust Gravity
in Flat FLRW Cosmology},
Zenodo preprint (2026),
\url{https://zenodo.org/records/23914715}.
\end{thebibliography}
\end{document}